\documentclass[a4paper,fleqn,usenatbib]{mnras}

\usepackage{newtxtext,newtxmath}

\usepackage[T1]{fontenc}
\usepackage{ae,aecompl}

\usepackage{graphicx}	
\graphicspath{{./Figures/}} 
\usepackage{grffile} 
\usepackage{amsmath}	
\usepackage{bm} 
\usepackage{subcaption} 
\usepackage{hyperref} 
\usepackage{xcolor} 
\usepackage{gensymb}
\usepackage{threeparttable} 

\newcommand{\Nh}{\ensuremath{\mathrm{N_{H_2}}}} 
\newcommand{\Cr}{\ensuremath{C_\mathrm{r}}} 

\title[Is Star Formation Driven By Column Density Alone?]{Spatial Statistics in Star Forming Regions:\\ Is Star Formation Driven By Column Density Alone?}

\author[B. Retter et al.]{
B. Retter,
J. Hatchell
and T. Naylor
\\
Physics and Astronomy, University of Exeter, Stocker Road, Exeter, EX4 4QL, UK}

\date{Accepted 2021 July 29. Received 2021 July 29; in original form 2020 July 23}

\pubyear{2021}

\begin{document}
\label{firstpage}
\pagerange{\pageref{firstpage}--\pageref{lastpage}}
\maketitle

\begin{abstract}
Star formation is known to occur more readily where more raw materials are available. 
This is often expressed by a `Kennicutt--Schmidt' relation where the surface density of Young Stellar Objects (YSOs) is proportional to column density to some power, $\mu$. 
The aim of this work was to determine if column density alone is sufficient to explain the locations of Class~0/I YSOs within Serpens South, Serpens Core, Ophiuchus, NGC1333 and IC348, or if there is clumping or avoidance that would point to additional influences on the star formation. 
Using the O-ring test as a summary statistic, 95 per cent confidence envelopes were produced for different values of $\mu$ from probability models made using the \textit{Herschel} column density maps.
The YSOs were tested against four distribution models: the best-estimate of $\mu$ for the region, $\mu=0$ above a minimum column density threshold and zero probability elsewhere, $\mu=1$, and the power-law that best represents the five regions as a collective, $\mu=2.05 \pm 0.20$. 
Results showed that $\mu=2.05$ model was consistent with the majority of regions and, for those regions, the spatial distribution of YSOs at a given column density is consistent with being random.
Serpens South and NGC1333 rejected the $\mu = 2.05$ model on small scales of $\sim 0.15~\mathrm{pc}$ which implies that small-scale interactions may be necessary to improve the model. 
On scales above 0.15~pc, the positions of YSOs in all five regions can be well described using column density alone.

\end{abstract}




\section{INTRODUCTION}
\label{sec:Intro}

Star formation is known to occur within molecular clouds and it has been found that star formation occurs more readily in regions of greater gas column density where more material is available for star formation. 
This effect can be observed on larger scales such as giant molecular clouds and galaxies where Kennicutt--Schmidt (K--S) relations show a higher surface density of star formation rate at higher column densities \citep{Kennicutt1989}.
It is observable on molecular cloud scales with similar star formation surface density relations \citep{2010ApJ...723.1019H}, and finally, smaller, filamentary scales where prestellar cores are more frequently observed coincident on the sky with high-density filaments \citep{andre2010}.
Since star formation surface density is correlated with column density, is it possible to describe the distribution of where star formation has occurred using only column density information?

The answer to this question requires a quantification of the amount that star formation is enhanced by increasing column density as well as a model for how this leads to star formation being distributed throughout a cloud. 
As mentioned earlier, measurements of the surface density of star formation rate as a function of column density are often expressed in the form of a power law (the K--S law is a particular form of this),
\begin{equation}
\label{eqn:intro-sigma}
\Sigma_\mathrm{SFR}= \Cr \Sigma^\mu_{\mathrm{Gas}},
\end{equation}
where $\Sigma_\mathrm{SFR}$ is the star formation rate surface density and $\Sigma_{\mathrm{Gas}}$ is the gas surface density.
The parameters $\Cr$ and $\mu$ quantify the relation between the surface densities of star formation and gas.\footnote{A full list of the mathematical symbols used in this paper is given in Table \ref{table:symbols}.}
Table \ref{table:power-law} presents some previous measurements of $\mu$ in local star forming regions.
Most $\mu$ values typically range between 1.5 and 2.5 \citep{2010ApJ...723.1019H,gutermuth2011,lombardi2014,Pokhrel2020} with some more extreme values of $\mu > 3$ \citep{lada2017}.

\begin{figure}
\includegraphics[width = \columnwidth]{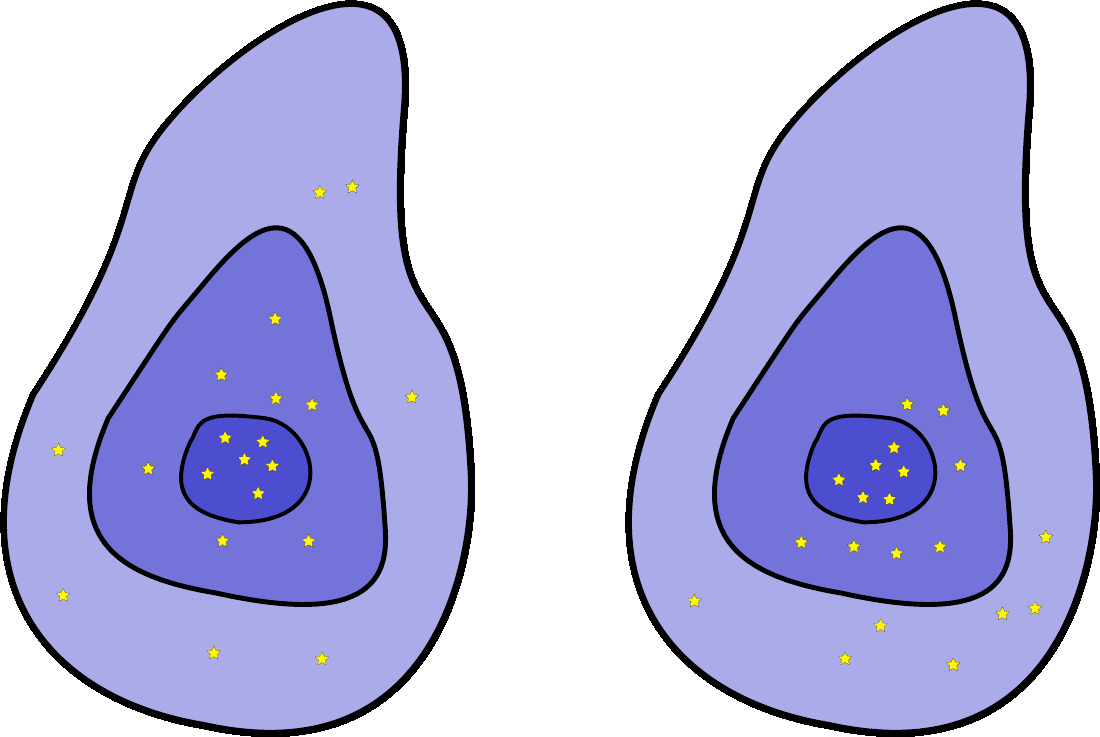}
\caption{\label{fig:intro_illustration} Illustration of two populations of YSOs with the same power-law relationship with column density.
The YSOs in the cloud on the left are evenly distributed throughout the cloud while those on the right are clustered towards the lower-right portion of the cloud. }
\end{figure}

A value of $\mu$ describes the change in the surface density of star formation as a function of column density, but it is missing information as to how the star formation is distributed throughout the cloud.
To illustrate this point Fig. \ref{fig:intro_illustration} presents two distributions of early-stage Young Stellar Objects (YSOs) which share the same value of $\mu$. 
The YSOs in the left-hand cloud are evenly distributed throughout the cloud according to column density, while those on the right are biased towards the lower-right of the region.
A distribution such as that in the right-hand cloud could be due to a column-density-independent effect which has strongly influenced the distribution of star formation, such as magnetic fields or stellar feedback. 
If such effects are significant, column density alone may not be sufficient to describe the distribution of star formation on local scales.

Using spatial statistics it is possible to investigate if the distribution of Class~0/I YSOs within a region is consistent with some distribution model, for example Eqn. \ref{eqn:intro-sigma} for given values of $\mu$ and $\Cr$.
These distribution models, known in spatial statistics as spatial point processes, are stochastic mechanisms for generating the locations of objects within a region.
To understand star formation, this work uses the locations of Class~0/I YSOs as a tracer for recent star forming activity to test how well different spatial point processes represent star formation within the molecular clouds.
Spatial statistics are well-suited for testing distribution models such as these -- examples of their application can be found in various fields including ecology and epidemiology \citep{Barot1999,Wiegand2009,Velazquez2016}. 

The O-ring summary statistic is a measure of how separated YSOs are from one another as a function of spatial scale.
This paper utilises this statistic to determine if the locations of a set of YSOs are consistent with a null hypothesis by comparing its values to confidence envelopes.
The spatial statistics methods used in this work for testing a model will be discussed in more detail in Section \ref{sec:spatial statistics}. 

In an earlier paper the efficacy of different summary statistics from spatial statistics were tested and then applied to Class~0/I YSOs in Serpens South to determine if they were consistent with complete spatial randomness (CSR) \citep{retter2019}.
CSR is a homogeneous Poisson point process where the probability of forming a star at a given location is unaffected by environment or neighbouring stars, and is therefore random.
The Class~0/I YSOs in Serpens South were found to be inconsistent with CSR and the work presented in this paper expands on this result by testing inhomogeneous Poisson Point Processes where the probability of placing a star at a given location is affected by the local column density.
In doing so it will be possible to determine if Class~0/I YSO positions, and therefore locations of star formation, are equivalent to randomly sampling a two-dimensional probability distribution based on the observed gas density.

While spatial statistics are useful for testing models, we employ Bayesian statistics for parameter fitting.
For this reason, this work introduces a Bayesian method of estimating $\mu$ by measuring the surface density of Class 0 and Class 1 Young Stellar Objects (YSOs) within column density bins in Section \ref{sec:bayes_stats}.

Class~0/I YSOs are used in this work as a tracer of star formation as they are the in the earliest stage of YSOs evolution. 
This can be observed in the relative distributions of YSOs where the Class 0 and Class I YSOs, whose positions tend to be more correlated with dense cloud than the more evolved Flat and Class~II and Class~III sources \citep{mairs2016,anne2020}.
Prestellar cores would likely be good tracers for star formation, but unfortunately it is not easy to determine which starless cores are likely to evolve to become stars and so Class~0/I YSOs are the youngest objects that are identifiable as definite precursors to stars.
This work will look at the YSOs in five local star forming regions which are good laboratories for testing distribution functions, due to their proximity and number of YSOs: Serpens South, Serpens Core, Ophiuchus, NGC1333 and IC348.

By applying the Bayesian method described in Section \ref{sec:bayes_stats}, the parameter $\mu$ was measured for each region individually as well as for the set of regions as a whole to estimate a global value of $\mu$ (Section \ref{subsec:mu}).
In addition to measuring $\mu$ for each region, we apply the same Bayesian method to find the best-estimates of $\Cr$ for each region in Section \ref{subsec:cr_estimate}.

In Section \ref{subsec:application} we use these best-fitting values for $\mu$ and $\Cr$ in combination with the O-ring statistic to test the distributions of YSOs in each of these regions against Eqn. \ref{eqn:intro-sigma}.
We also present the results of testing the distributions of YSOs for general models with $\mu = 0$, $\mu = 1$ and the global value of $\mu = 2.05$.
Each of these models test for a different potential physical description for how the distribution of stars is correlated with column density. 
The first model uses $\mu=0$ above a threshold visual extinction $\mathrm{A_v} = 6$ to test if there is a relationship between the amount of column density and surface density of YSOs, or if, once some threshold is reached, the YSOs are simply dispersed randomly.
The second model, $\mu=1$, is motivated by the distribution of prestellar cores which may depend less steeply on column density than protostellar cores \citep{Sokol2019,konyves2020}.
With this model we look to determine if the observed distribution of YSOs is consistent with the distribution they may have had at an earlier stage in their evolution.
The third model is to explore how well, or even if, a single power-law, can simultaneously represent multiple star-forming regions. 

It was found that, when considering the number of regions that reject the model, the region-specific $\mu$ values were the most successful at describing the distributions of Class~0/I YSOs. 
As for the general models, the best-performing model was the global model of $\mu = 2.05$ which was consistent with 3 out of the 5 regions it was applied to, and the worst performing model was $\mu=1$ -- which was rejected by every region.
As a further test, we apply the $\mu = 2.05$ model to the Class~II YSOs in all five regions in Section \ref{subsec:results-classII}.
This is to show that these statistics have enough discriminatory power to distinguish between two potentially similar distributions within the same study region. 

Finally, the discussion of these results is presented in Section \ref{sec:discussion}.
There we discuss how effective a column-density-only model is at describing the distributions of YSOs.
We also explore how the rejection of the $\mu = 1$ model by Class~0/I YSOs could imply an environmental dependence on the evolutionary time-scales for prestellar cores and/or Class~0/I YSOs. 

\begin{table*}
\centering
\begin{threeparttable}
\caption{\label{table:power-law} YSO Kennicutt-Schmidt power law estimates for different regions.}
\begin{tabular}{l l l }
Region                            & Power-law    & Source               \\
\hline
AFGL 490                          & $1.8 \pm 0.3$   & \citep{Pokhrel2020}   \\
Aquila North                      & $1.8 \pm 0.1$   & \citep{Pokhrel2020}   \\
Aquila South                      & $2.3 \pm 0.2$   & \citep{Pokhrel2020}   \\
Auriga-California Molecular Cloud & 4            & \citep{harvey2013}    \\
California                        & $3.31 \pm 0.23$ & \citep{lada2017}      \\
Cep 0B3                           & $2.2 \pm 0.1$   & \citep{Pokhrel2020}   \\
Cygnus X                          & $1.9 \pm 0.1$   & \citep{Pokhrel2020}   \\
G305                              & $2.50 \pm 0.04$ & \citep{willis2015}    \\
G326.4                            & $1.91 \pm 0.05$ & \citep{willis2015}    \\
G326.6                            & $1.77 \pm 0.04$ & \citep{willis2015}    \\
G333                              & $2.86 \pm 0.03$ & \citep{willis2015}    \\
G351                              & $2.30 \pm 0.03$ & \citep{willis2015}    \\
Mon R2                            & $2.1 \pm 0.1$   & \citep{Pokhrel2020}   \\
MonR2                             & $2.67 \pm 0.02$ & \citep{gutermuth2011} \\
NGC 2264                          & $1.8 \pm 0.1$   & \citep{Pokhrel2020}   \\
NGC 6634                          & $2.08 \pm 0.08$ & \citep{willis2015}    \\
Ophiuchus                         & $1.87 \pm 0.03$  & \citep{gutermuth2011} \\
Ophiuchus                         & $1.9 \pm 0.1$   & \citep{Pokhrel2020}   \\
Orion A                           & $1.99 \pm 0.05$ & \citep{lombardi2013}  \\
Orion A                           & $2.2 \pm 0.1$   & \citep{Pokhrel2020}   \\
Orion B                           & $2.16 \pm 0.10$ & \citep{lombardi2014}  \\
Orion B                           & $2.3 \pm 0.2$   & \citep{Pokhrel2020}   \\
Perseus                           & $2.4 \pm 0.6$   & \citep{zari2016}      \\
Perseus                           & $2.1 \pm 0.1$   & \citep{Pokhrel2020}   \\
S140                              & $1.8 \pm 0.2$   & \citep{Pokhrel2020}  \\
\hline
\end{tabular}
\begin{tablenotes}
\item Note: Different sources use different methods and different astrophysical objects to produce power-law estimates. 
\end{tablenotes}
\end{threeparttable}
\vspace{-4mm}
\end{table*}

\section{SPATIAL STATISTICS}
\label{sec:spatial statistics}
To determine if there is a correlation between the gas density within a local molecular cloud and the number of Class~0/I YSOs it is necessary to produce and test a model that encompasses this relationship. 
The type of model that will be described in this paper is known in the field of spatial statistics as a spatial point process, which is a stochastic mechanism that generates a set of points within a study region with the number and locations of these points dependent on the model. 
A spatial point pattern is a realisation of the model, and by assuming YSO positions are a realisation of a model it is possible to try and infer information about the model from the observed pattern.
As such, these models and the realisations thereof are informed by physics but are not physical simulations.

\subsection{First-order models}
First-order effects affect the expected number of stars at a given location without consideration of the presence, or absence, of other stars.
Due to this lack of dependence on neighbouring stars these are sometimes attributed to environmental factors. 
For example the availability of material for forming stars could be a factor in the number and location of stars formed within a star-forming region. 
One measure of first-order effects, the first-order intensity, is the expected density at a given position $x$.
The first-order intensity is given by the equation \citep{Diggle2013}
\begin{equation}
\label{eqn:first-order}
\lambda(x) = \lim_{|dx|\rightarrow 0} \left\{\frac{\mathrm{E}[N(dx)]}{|dx|}\right\},
\end{equation}
where $\lambda(x)$ is the first-order intensity at position $x$, $dx$ is the small region containing $x$, $\mathrm{E}[N(dx)]$ is the expected number of points contained within $dx$ and $|dx|$ denotes the area within $dx$.
Given the first-order intensity it is then possible to estimate the mean number of stars within the study window with
\begin{equation}
\langle N\rangle=\int_A\lambda(x)dx
\label{eqn:pois_nbar}
\end{equation}
where $N$ is the total number of stars within study window $A$. 

\subsection{Generating spatial point patterns from first-order models}
\label{subsec:patterns}
With first-order models the probability of placing a star is entirely defined by the first-order intensity, and so a spatial point pattern can be produced by randomly sampling the first-order intensity distribution across the study region.

\begin{figure*}
\includegraphics[width = \textwidth]{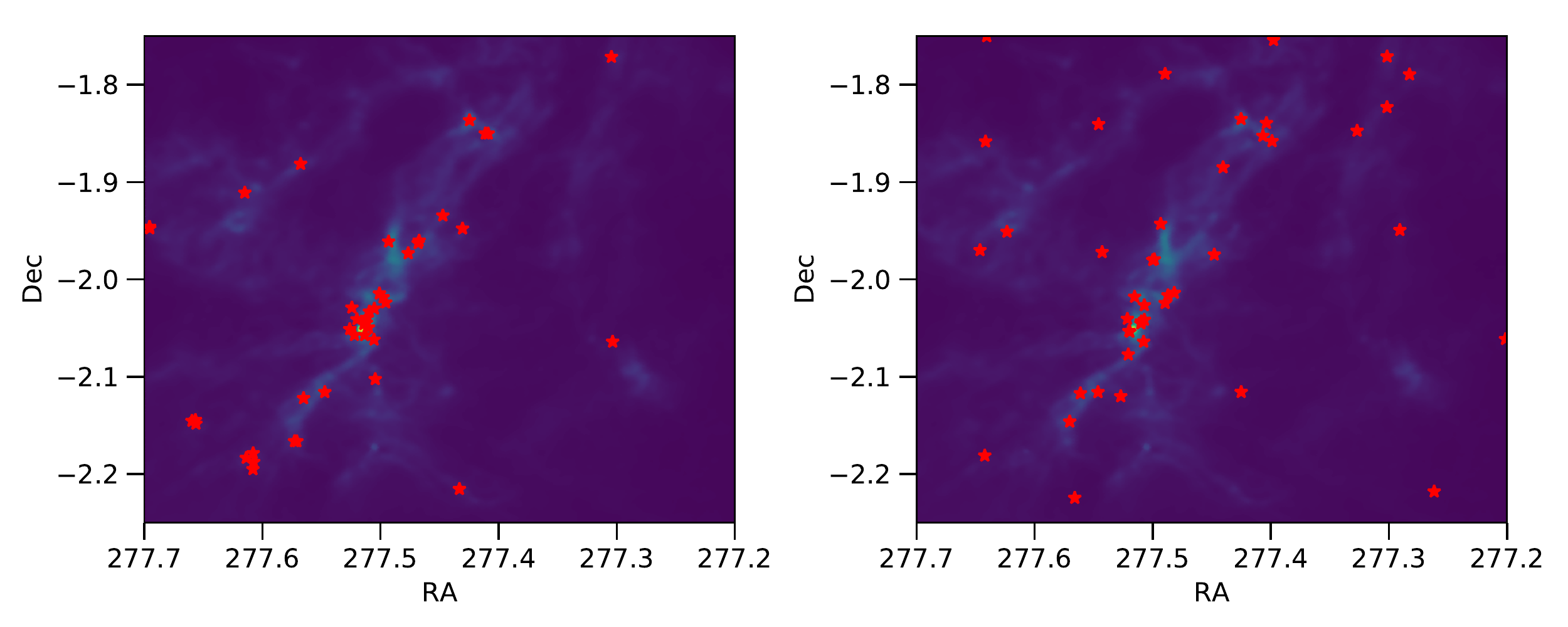}
\caption{\label{fig:serpens_south_random} (left) Class~0/I YSOs in Serpens South and (right) random realisation of YSOs with $\mu = 2.05$ (see Section \ref{subsubsec:results-mu2}) plotted on \textit{Herschel} 18.2$''$ column density maps.}
\end{figure*}

For a study region represented by a grid some modification needs to be made to account for cells that cover different amounts of area.
The joint-distribution is then composed of two parts: a probability proportional to area and a probability proportional to the first-order intensity. 
Therefore, when making a pattern using a first-order model in a gridded study region we sample
\begin{equation}
\label{eqn:prob_u}
\mathrm{prob}(u) =  \frac{\lambda(u) |u|}{\sum_v \lambda(v) |v|},
\end{equation}
where $u$ and $v$ are cell indices and $|u|$ and $|v|$ are cell areas.
The left panel of Fig. \ref{fig:serpens_south_random} shows the positions of the Class~0/I YSOs in Serpens South with an example realisation of Eqn. \ref{eqn:prob_u} with $\lambda(u)$ equal to Eqn. \ref{eqn:intro-sigma} using $\mu = 2.05$ in the right panel.
Given that they were generated using Eqn. \ref{eqn:intro-sigma}, it is known that the YSOs in the right panel are consistent with this spatial point process and the question remains as to whether or not the YSOs in the left panel are also consistent with this process.

\subsection{Spherical Projection}
\label{subsec:spherical}
The column density data used in this paper are sections of the celestial sphere which have been stored in a tangent plane projection.
To calculate the O-ring statistic, estimate the quantity $\mu$ using the methodology in Section \ref{sec:bayes_stats} and accurately reproduce first-order spatial point processes the distances between YSOs and the sizes of pixels need to be measured.
Both of these quantities are affected by the gnomonic projection.

The projected maps are stored in the FITS file format \citep{1981A&AS...44..363W} which contains the keywords necessary to convert pixel coordinates to coordinates in the gnomonic projection plane or celestial sphere.
The set of keywords that allow for coordinate conversion are known as the FITS ``world coordinate system'' (WCS) and with these the pixel coordinates of most maps can be converted to projection and celestial coordinates \citep{calabretta_celestial,greisen_world}.
Software is available to perform transform between these coordinates systems; the results in this paper use the library {\scshape wcslib} \footnote {https://www.atnf.csiro.au/people/mcalabre/WCS/wcslib/}. 

While it is possible to use projection coordinates entirely, they are representations of sections of a spherical surface and so it may be more simple and intuitive to use the native spherical coordinates.
For example, with spherical coordinates the angular distance, $\Delta\sigma$, between two points can be calculated using the haversine formula,
\begin{equation}
\label{eqn:great_circle}
\Delta \sigma = 2\mathrm{arcsin}\sqrt{\mathrm{sin}^2\left(\frac{\Delta \delta}{2}\right)+\mathrm{cos}\delta_1\mathrm{cos}\delta_2\mathrm{sin}^2\left(\frac{\Delta \alpha}{2}\right)}
\end{equation}
where $\alpha$ and $\delta$ refer to the right ascension (RA) and declination (Dec) in radians. 

Some projections distort the shape or size of areas on the sphere and so the amount of angular area represented by a given pixel on the map is not necessarily consistent across the projection plane. 
The \textit{Herschel} maps used in Section \ref{sec:results}, use the gnomonic projection which is designated the AIPS code `TAN'.
A gnomonic projection is produced by projecting points on the surface of a sphere onto a tangential plane from the perspective of an observer at the sphere's centre.
With this type of projection the amount of distortion increases as a function of angular distance measured from the tangent point, $\theta$.
This distortion increases the projected area covered by objects on the sphere by $1/\mathrm{cos}^3(\theta)$ and, therefore, decreases the amount of area  on the sphere covered by a pixel in the projection by $\mathrm{cos}^3(\theta)$.
Finally, the amount of area covered by a pixel is given by,
\begin{equation}
|u| = |O|\mathrm{cos}^3(\theta_{u}),
\end{equation}
where $|u|$ is the area covered by pixel $u$, $|O|$ is the area of a pixel at the tangent point. The angle $\theta_u$ is equal to $\Delta \sigma$ between the tangent point and $u$.
Unlike angular distance, which requires only the angular coordinates for two points, the area covered by a pixel depends on the type of projection. 
As a general solution, however, it is possible to calculate the world coordinates of the corners of each pixel and use those to approximate the areas, though this assumes the pixels are small enough to be approximately flat.

\subsection{Hypothesis testing}
\label{subsec:spatial-hypothesis}

It is possible to test whether a distribution of stars is consistent with a given spatial point process through the use of summary statistics and confidence envelopes.
The most common spatial point process that patterns are tested against is that of complete spatial randomness (CSR) as in \citet{retter2019}, but other processes can be tested against using the methods described in this section.

\subsubsection{Summary statistic}
\begin{figure}
\includegraphics[width=\columnwidth]{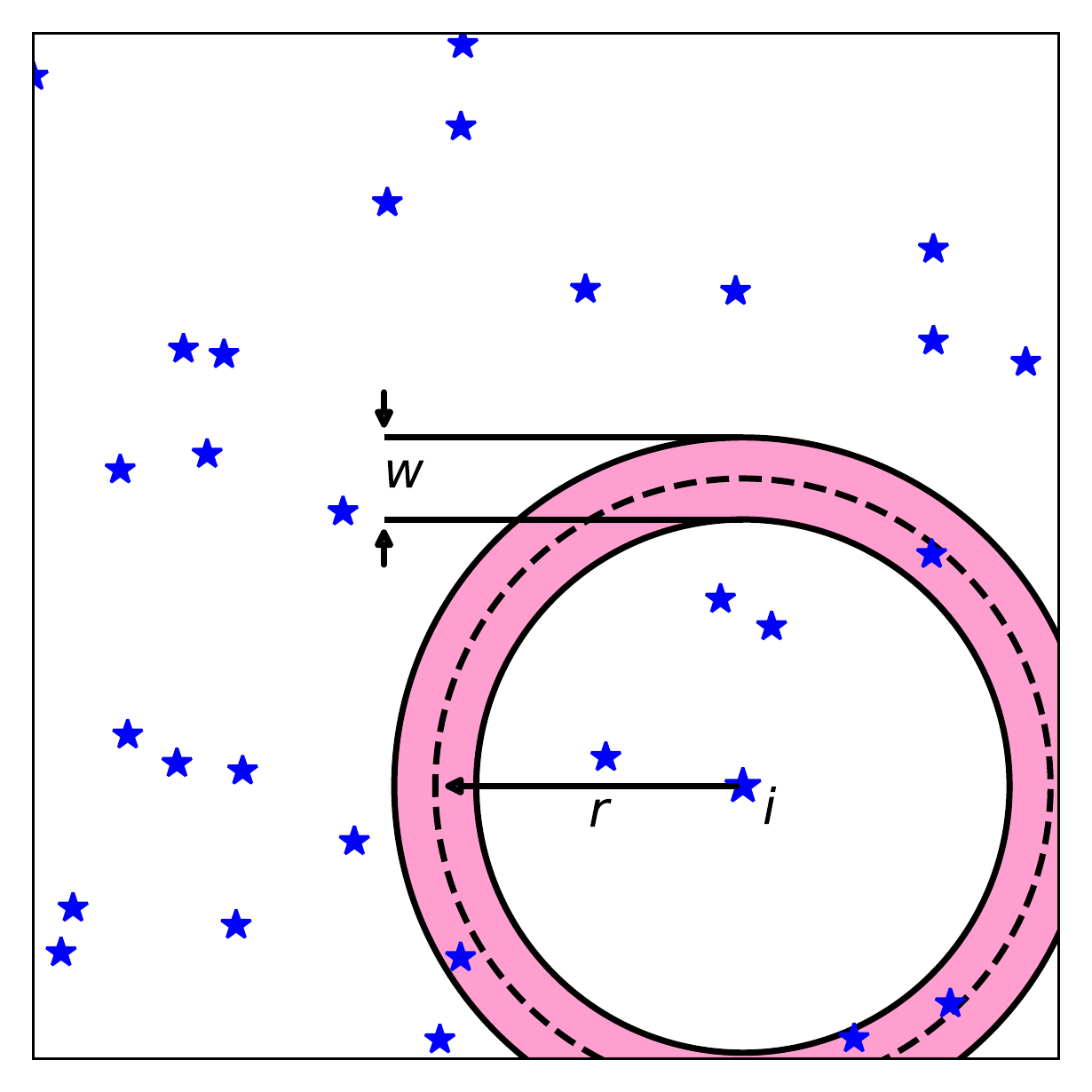}
\caption{\label{fig:spatial_statistics:Oring_explainer} Schematic illustration of O-ring function.  Density at a distance $r$ from star $i$ is estimated by counting stars within the shaded portion. The shaded area shows the region where $\mathrm{S}(u,v)$ and $\mathrm{I}_r(x_{u,v},y_{u,v},x_i,y_i)$ are equal to 1.}
\end{figure}

The summary statistic that will be used and discussed in this paper is the O-ring statistic.
O-ring uses all of the inter-point distances to estimate the average density of stars that would be observed at a distance $r$ from a given star.
As is the case in this work, study windows may be in the form of a grid, and to account for this the methodology of \citet{Wiegand2004} is applied to calculate O-ring. 
In general, O-ring can be calculated using
\begin{equation}
\label{eqn:oring}
\mathrm{\hat{O}}(r) = \frac{\sum_{i=1}^{N}\text{Points}[\mathrm{R}_i^w(r)]}{\sum_{i=1}^{N}\text{Area}[\mathrm{R}_i^w(r)]},
\end{equation}
where $\mathrm{R}_i^w(r)$ is an annulus with radius $r$ and width $w$ centred on the $i$th star, and the operators $\text{Points}[\mathrm{R}_i^w(r)]$ and $\text{Area}[\mathrm{R}_i^w(r)]$ count the number of stars and area contained within $\mathrm{R}_i^w(r)$ respectively.
Fig. \ref{fig:spatial_statistics:Oring_explainer} shows a schematic representation of the O-ring function where stars and area within the shaded annulus of Fig. \ref{fig:spatial_statistics:Oring_explainer} are counted for star $i$. 
If boundary conditions are applied then only the points and area within the intersection of $\mathrm{R}_i^w(r)$ and the study region are counted.

The number of points within $\mathrm{R}_i^w(r)$ in a grid is defined by
\begin{equation}
\text{Points}[\mathrm{R}_i^w(r)] = \sum_u \sum_v \mathrm{S}(u,v)\mathrm{P}(u,v)\mathrm{I}_r(x_{u,v},y_{u,v},x_i,y_i).
\end{equation}
Here, $u$ and $v$ are the row and column indices of the grid respectively. 
$\mathrm{S}(u,v)$ is an indicator function equal to 1 if cell $(u,v)$ is contained within the study window and zero otherwise and $\mathrm{P}(u,v)$ is the number of stars contained within $(u,v)$.
Finally, $\mathrm{I}_r$ is another selection function to determine if a cell is within the annulus, defined by
\begin{align}
\label{eqn:points_operator}
\mathrm{I}_r(x_{u,v},y_{u,v},x_i,y_i) = \begin{cases} 
1\ &\text{ if } r-\frac{w}{2} \leq d_{(u,v),i} \leq r + \frac{w}{2},\\
0\ &\text{otherwise,}\\
\end{cases}
\end{align}
where $d_{(u,v),i}$ is the the distance between the centre of cell $(u,v)$ and the $i$th star. The $\mathrm{Area}$ operator within Eqn. \ref{eqn:oring} is similarly defined
\begin{equation}
\label{eqn:area_operator}
\text{Area}[\mathrm{R}_i^w(r)] = \sum_u \sum_v \mathrm{S}(u,v)\mathrm{A}(u,v)\mathrm{I}_r(x_{u,v},y_{u,v},x_i,y_i),
\end{equation}
where $\mathrm{A}(u,v)$ is the amount of area contained within cell $(u,v)$. 
There exist other summary statistics such as Ripley's K \citep{Ripley1981}, Diggle's G function \citep{Diggle2013} and the "free-space" function; however, the O-ring test has been chosen for this analysis due to its sensitivity compared to the other tests \citep{retter2019}.

\subsubsection{Confidence envelopes}
A summary statistic produces a single value (in this case at a given $r$) to represent a chosen facet of the data being measured. 
This value can be compared to the distribution of values the summary statistic takes under some null hypothesis, $H_0$ to determine if the null hypothesis can be rejected with some significance level, $\alpha$.
The distribution of a summary statistic under a simple null hypotheses, such as that of CSR, can sometimes be found analytically \citep{wiegand2016}; otherwise it may be sampled computationally through repeated realisations of the $H_0$. 
The global confidence envelope covers the range of acceptable values of the summary statistic which, if it is exceeded, rejects $H_0$ with significance $\alpha$.

To test the significance of a single measurement it is sufficient to find the distribution of the statistic under the null hypothesis and determine if the measured value is among the $k$th-most-extreme values where, for a two-sided distribution, 
\begin{equation}
\label{eqn:alpha}
\alpha = 2k/(n+1)
\end{equation}
and $n$ is the number of simulated patterns of $H_0$. 

Where there are multiple measurements being tested simultaneously, using the $k$th-most-extreme values for each results in a lower $\alpha$ for the test as the probability of finding an extreme values increases as the number of independent tests increases. 
It is possible to test over multiple scales with a controlled $\alpha$ by using a global confidence envelope.

A global confidence envelope is able to identify significant measurements of an observed statistic, $T_1(r)$. 
Using the methodology described in \citet{myllmaki2017} the upper and lower bounds of the global confidence envelope are defined by
\begin{align}
\label{eqn:Tlow}
T^u_{\mathrm{low}}(r) &= T_0(r) - u_{\alpha}|\underline{T}(r) - T_0(r)|\\
\label{eqn:Tupp}
T^u_{\mathrm{upp}}(r) &= T_0(r) - u_{\alpha}|\overline{T}(r) - T_0(r)| 
\end{align}
respectively, where $T_0(r)$ is the expected value under the $H_0$, $\overline{T}(r)$ and $\underline{T}(r)$ are the 2.5 per cent upper and lower quantiles of the distribution of $T(r)$ under $H_0$ and $u_{\alpha}$ is the parameter used to determine the confidence level of the envelopes. 
If the distribution of $T(r)$ is not known then it can be estimated from measurements of simulated patterns $T_i(r)$, where $i = 2,3,\  \ldots\ , n+1$.
In this paper $u_{\alpha}$ is the $\alpha (n+1)$th largest value from the collection of $u_i$s calculated using
\begin{equation}
u_i = \max \left[f(r)^{-1} (T_i(r)-T_0(r)) \right]
\end{equation}
where $f(r)$ is 
\begin{align}
f(r) = \begin{cases} 
\overline{T}(r)-T_0(r)\ &\text{ if } \{T(r) \geq T_0(r)\} \text{, or}\\
\underline{T}(r)-T_0(r)\ &\text{ if } \{T(r) < T_0(r)\}. \\
\end{cases}
\end{align}
Using Eqns. \ref{eqn:Tlow} and \ref{eqn:Tupp} it is possible reject a null hypothesis with a controlled $\alpha$ and find the values of $r$ which exceed the envelope.

\section{BAYESIAN STATISTICS}
\label{sec:bayes_stats}

\begin{figure*}
\includegraphics[width=\linewidth]{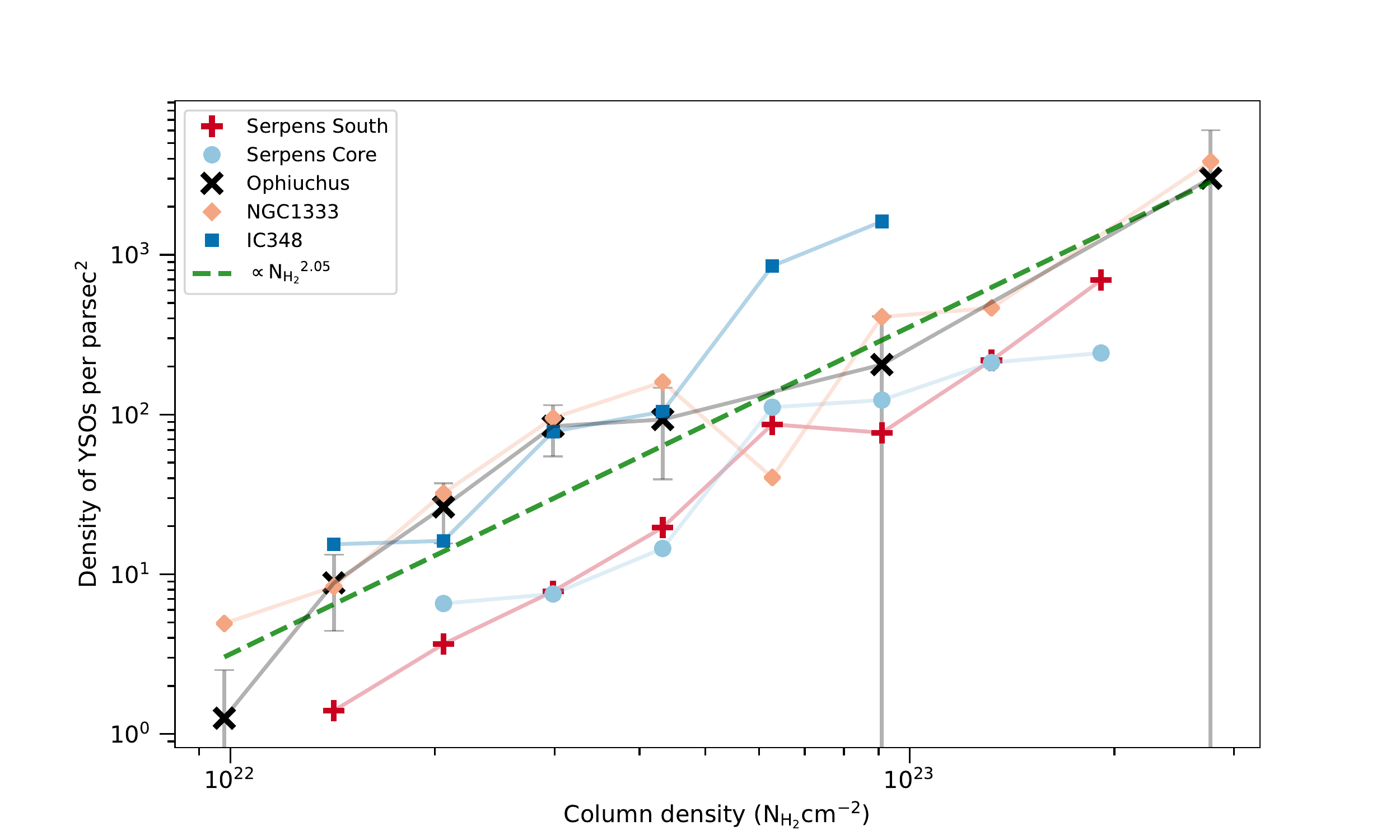}
\caption{\label{fig:results-yso_surface_density} YSO surface density measurements within column density bins in Serpens South, Serpens Core, Ophiuchus, NGC1333 and IC348. 
As an illustration of typical uncertainties, the Poisson uncertainties on YSO counts are given. 
The dashed line shows a gradient of 2.05 (Section \ref{subsec:mu}).}
\end{figure*}

From inspection of Eqn. \ref{eqn:intro-sigma}, the power-law index $\mu$, for a single region, can be estimated from the straight-line gradient in a plot of $\log(\Sigma_\mathrm{SFR})$ versus $\log(\Sigma_\mathrm{Gas})$ \citep{2010ApJ...723.1019H,gutermuth2011}.
A plot of the YSO surface density versus column density for the five regions studied in this paper is presented in Fig. \ref{fig:results-yso_surface_density} and from this it is clear that there is a correlation between YSO surface density and column density in each of these regions.
The gradients ($\mu$ values) are not dissimilar, but different $\Cr$ values are implied as the YSO surface density functions have different y-intercepts for different regions.
One problem with measuring gradients is that the number of YSOs in each $\mathrm{A_v}$ bin are small enough that care needs to be taken when dealing with the uncertainties.
Care must also be taken when estimating $\mu$ values that represent multiple regions as combining area and YSO counts to find an average density assumes every region has the same value of $\Cr$, which is not always true.
For these reasons a Bayesian method of estimating $\mu$ is described in this section which can be extended to calculate joint values of $\mu$ for sets of regions.

To model the distribution of Class~0/I YSOs with respect to the observed column density in a star forming region we introduce the following equation,
\begin{equation}
\hat{\lambda}\mathrm{(\Nh)} = \Cr\times\Nh^\mu,
\label{eqn:lmda-model}
\end{equation}
where $\hat{\lambda}\mathrm{(\Nh)}$ is the estimate of the number density of Class~0/I YSOs at column density $\Nh$, $\Cr$ is a region-specific constant that normalises the number of Class~0/I YSOs such that Eqn. \ref{eqn:pois_nbar} is satisfied for a given region, and $\mu$ is the global power law affecting the distribution of YSOs with respect to the column density.
We then look to find the most likely values of $\Cr$ and $\mu$. 

First, consider a sub-region, $m$, of the star-forming region that has an approximately constant column density, $\Nh{_{,m}}$. 
If we assume the probability that this sub-region contains a number of YSOs, $N_m$, follows a Poisson distribution, it can then be shown that the probability of counting $N_m$ YSOs is given by
\begin{multline}
\mathrm{prob}(N_m|A_m,\Nh{_{,m}},\Cr,\mu) =\\
 \frac{(\Cr(\mathrm{\Nh{_{,m}})^{\mu}}A_m)^{N_m}e^{-(\Cr\mathrm{(\Nh{_{,m})}^{\mu}} A_m)}}{N_m!},
\end{multline}
where $A$ is the area of $m$.
Repeating the experiment with $M$ different sub-regions of the star-forming region results in the probability
\begin{multline}
\label{eqn:N_one_region}
\mathrm{prob}(\{N\}|\{A\},\{\Nh\},\Cr,\mu) =\\
 \prod_{m=0}^M\frac{(\Cr{(\Nh{_{,m}}})^{\mu}A_m)^{N_m}e^{-(\Cr{(\Nh{_{,m}})}^{\mu}A_m)}}{N_m!},
\end{multline}
where $\{N\}$ is a vector of $N$, etc.
Given the condition of similar column densities, this equation functions with any form of subdivision of the star-forming region. 
For this paper contours of column density are used to define each sub-region.

If we assume a uniform prior for $\mu$ and a Jeffreys' prior for $\Cr$, we find that the na\"ive prior of $\mu$ and $\Cr$ is inversely proportional to $\Cr$, i.e.
\begin{align}
\label{eqn:one-region-prior}
\mathrm{prob}(\Cr,\mu) = \begin{cases} 
\frac{1}{\Cr}\ &\text{ for } \Cr \geq 0 \text{ and } \mu \geq 0,\\
0\ &\text{otherwise.}\\
\end{cases}
\end{align}
With Bayes' theorem we may then construct an equation that can be used to find the probability associated with a combination of $\Cr$ and $\mu$:
\begin{multline}
\label{eqn:c,b_one_region}
\mathrm{prob}(\Cr,\mu|\{N\},\{A\},\{\Nh\}) \propto \\
 \frac{1}{\Cr}\prod_{m=0}^M\frac{(\Cr{(\Nh{_{,m}}})^{\mu}A_m)^{N_m}e^{-(\Cr{(\Nh{_{,m}})}^{\mu}A_m)}}{N_m!}.
\end{multline}
This joint Probability Density Function (PDF) can be marginalised to find the marginal probabilities of $\mu$ and $\Cr$ for the star-forming region separately.

As an extension, we may then question if the power in Eqn. \ref{eqn:lmda-model} is not specific to a single star-forming region, but is instead a universal property shared across different star-forming regions, each with unique values of $\Cr$.
It would be desirable, then, to combine the measurements from multiple star-forming regions, each with their own sub-regions, to produce a single best-estimate for $\mu$. 
These measurements can be included by modifying the prior in Eqn. \ref{eqn:one-region-prior} to be proportional to the inverse of the product of $\Cr$ values,
\begin{equation}
\label{eqn:more-region-prior}
\mathrm{prob}(\{\Cr\},\mu) \propto \prod_{i=1} ^\mathit{\#\ s.f.\ regions} \frac{1}{{\Cr}_{,i}},
\end{equation}
and adding the measurements to the product in Eqn. \ref{eqn:c,b_one_region} to produce the equation  
\begin{multline}
\label{eqn:c,b_all_region}
\mathrm{prob}(\{\Cr\},\mu|\{N\},\{A\},\{\Nh\}) \propto \\
\prod_{i=1}^\mathit{\#\ s.f.\ regions} \frac{1}{{\Cr}_{,i}}\prod_{m=0}^M\frac{({\Cr}_{,i}{(\Nh{_{,m}}})^{\mu}A_m)^{N_m}e^{-({\Cr}_{,i}{(\Nh{_{,m}})}^{\mu}A_m)}}{N_m!}.
\end{multline}

The addition of more regions increases the number of dimensions of the PDF which increases the computational difficulty of sampling the PDF without the use of techniques such as Markov chain Monte Carlo (MCMC) \citep{hastings1970,2010CAMCS...5...65G,Foreman_Mackey_2013}.

\section{APPLICATION TO STAR-FORMING REGIONS }
\label{sec:results}

In this work $\mu$ values are determined for the star-forming regions Serpens South, Serpens Main, Ophiuchus, NGC1333 and IC348, as well as a joint $\mu$ value for all regions. 
The joint $\mu$ value and $\mu = 0$ and $\mu = 1$ were then tested using 95 per cent confidence envelopes as described in Section \ref{sec:spatial statistics}. 

The \citet{dunham2015} YSO catalogue was used for YSO position and classification for all regions.
YSOs classified as Class~0/I are those with a corrected spectral index value greater than or equal to 0.3 and $T_\text{bol} < 650 \mathrm{K}$. 
Using a single catalogue and classifying with corrected spectral index and $T_\text{bol}$ provides a simple and consistent method of identifying YSO populations which can be compared between clouds. Missing or misclassified sources should not affect the results unless there is a selection bias with respect to column density, which we do not expect except possibly at high column densities (discussed further in Sect.~\ref{sec:discussion}).
The column density data used for the different regions were the \textit{Herschel} 18.2$''$ resolution maps \citep{andre2010,palmeirim2013}.
Table \ref{table:region_summary} lists the number of Class~0/I YSOs and distances to each cloud, as well as the RA and Dec boundaries used to extract the regions.

\begin{table*}
    \begin{threeparttable}
    \caption{\label{table:region_summary} Summary of the properties of the star-forming regions analysed.}
    \begin{tabular}{c c c c c}
    Region & No. of & RA limits & Dec limits & Distance (pc) \\
    & Class~0/I & & &\\
    & YSOs & & &\\
    \hline
    Serpens South & 44 & $277.2\degree \leq \mathrm{RA} \leq 277.7\degree$ & $-2.25\degree \leq \mathrm{Dec} \leq -1.75\degree$ & 484 \tnote{a}\\
    Serpens Core & 16 & $277.4\degree \leq \mathrm{RA} \leq 277.6\degree$ & $1.18\degree \leq \mathrm{Dec} \leq 1.38\degree$ & 484 \tnote{a}\\
    Ophiuchus & 24 & $246.0\degree \leq \mathrm{RA} \leq 248.5\degree$ & $-25.2 \degree\leq \mathrm{Dec} \leq -23.8\degree$ & 144 \tnote{a}\\
    NGC1333	 & 32 & $52.0\degree \leq \mathrm{RA} \leq 52.8\degree$ & $31.0\degree \leq \mathrm{Dec} \leq 31.8\degree$ & 293 \tnote{b}\\
    IC348 & 12 & $55.8\degree \leq \mathrm{RA} \leq 56.4\degree$ & $31.9\degree \leq \mathrm{Dec} \leq 32.5\degree$ & 321 \tnote{b}\\
    \hline
    \end{tabular}
    \begin{tablenotes}
    \item[a] \citep{zucker2019}
    \item[b] \citep{Ortiz2018}
    \end{tablenotes}
    \end{threeparttable}
\end{table*}
\subsection{Estimations of $\mu$}
\label{subsec:mu}

\begin{figure}
\includegraphics[width=\columnwidth]{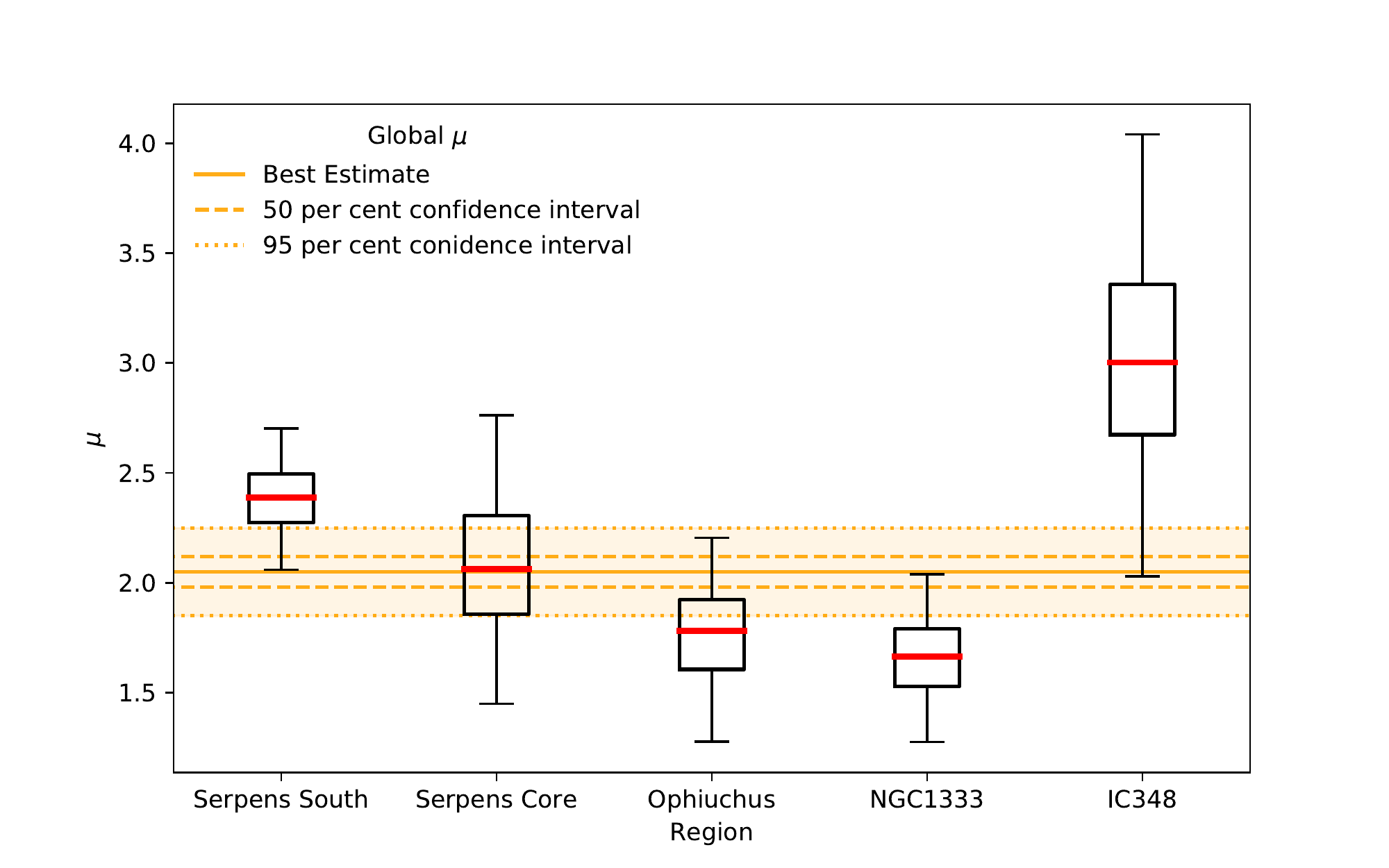}
\caption{\label{fig:mu_regions} Marginalised distributions for $\mu$ for Serpens South, Serpens Core, Ophiuchus, NGC1333, IC348 and global estimate. The box and whiskers present the best estimate and the 50 and 95 per cent intervals. The orange lines present the best estimate and the 50 and 95 percent intervals for the global $\mu$ value.}
\end{figure}

In this section, we present the results of estimating the power law using the Bayesian marginalisation described in Section \ref{sec:bayes_stats}.
The joint-probability distributions for $\Cr$ and $\mu$ were calculated using the number of YSOs and the area contained within column density bins in each region. The results of this analysis are presented in Fig. \ref{fig:join-pdf-all}.
These joint-probability distributions were then marginalised over $\Cr$ to find the PDFs for $\mu$ for each region, the results of which are shown in Fig. \ref{fig:mu_regions} and Table \ref{table:mu}.

To determine if these joint-probability distributions were reasonable Eqn. \ref{eqn:pois_nbar} was used to find the solutions where $\langle N\rangle$ was equal to the observed number of YSOs in each region.  
These are overlaid on Fig. \ref{fig:join-pdf-all} as dashed lines.
The intersection between Eqn. \ref{eqn:pois_nbar} and the high probability density regions of the joint probability distribution demonstrates that Eqn. \ref{eqn:c,b_one_region} is consistent with producing $\Cr$ and $\mu$ values that approximate the number of YSOs used to calculate the joint-distribution.
As can be seen in Fig.~\ref{fig:join-pdf-all}, this overlapping of the two functions is consistent across all regions.

To find the most likely value of $\mu$ over all regions the areas and number of YSOs in each sub-region were combined into a single 6-dimensional joint-probability distribution: one dimension for each of the five regions' $\Cr$ values and one for $\mu$.
This distribution was sampled and marginalised using the MCMC functionality of the {\sc Python} package {\scshape emcee} \citep{Foreman_Mackey_2013} to find the PDF for $\mu$.
The best-estimate for the global power law was found to be $\mu = 2.05 \pm 0.20$, using a 95 per cent confidence interval as the uncertainty.

A value of $\mu = 2.05$ sits within the 95 per cent confidence intervals for Serpens Core, Ophiuchus and IC348 and only marginally outside those of Serpens South and NGC1333. 
While further testing is required to determine if the distributions of the YSOs within these regions are consistent with a global power law, the global power law appears to adequately describe the distribution of these YSOs as a set.

{\graphicspath{{./Figures/beta_pdf/rescaled/}}
\begin{figure*}
	\begin{subfigure}[b]{0.45\linewidth}
		\includegraphics[width=1.1\linewidth]{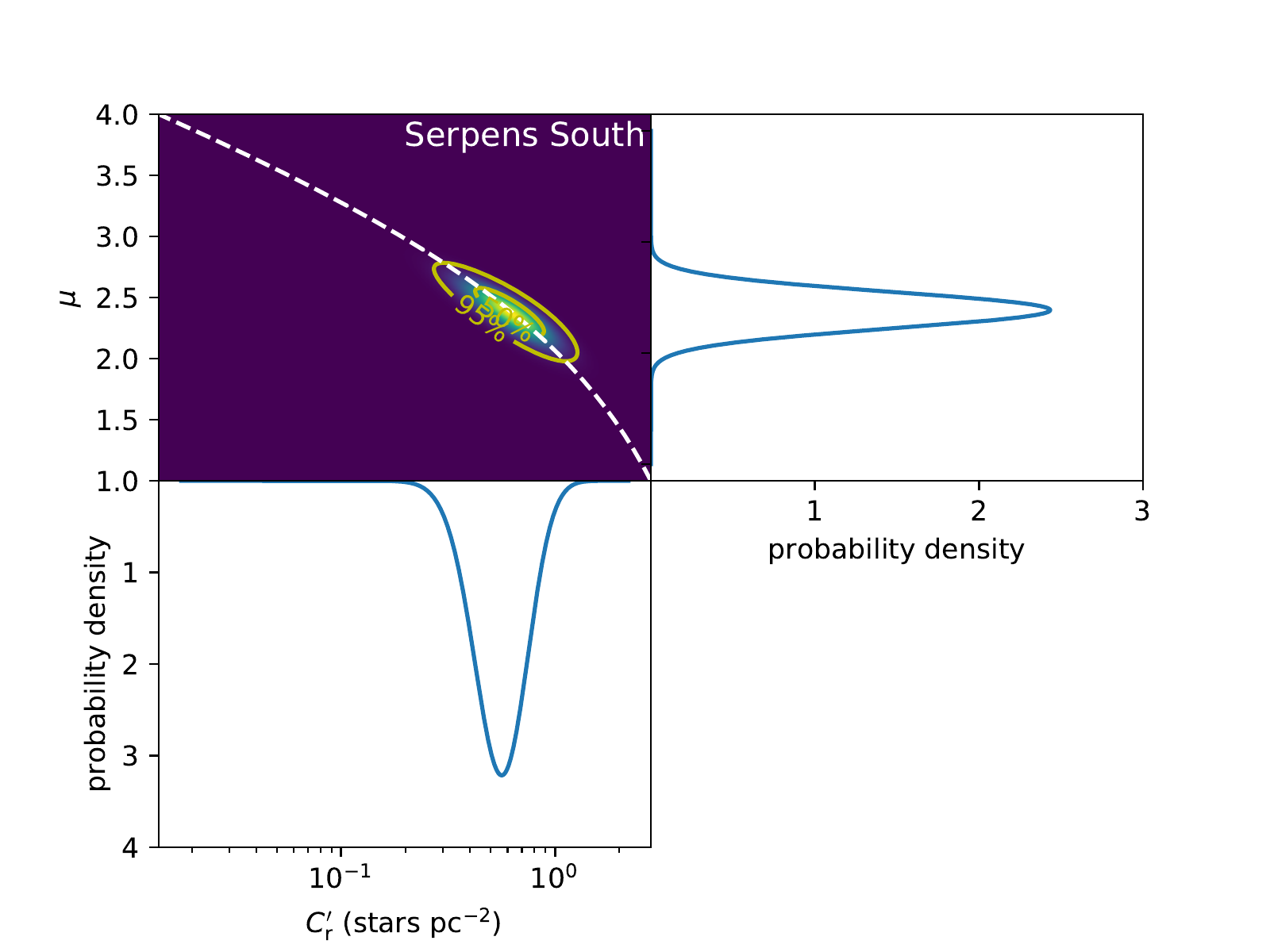}
	\end{subfigure}
	\begin{subfigure}[b]{0.45\linewidth}
		\includegraphics[width=1.1\linewidth]{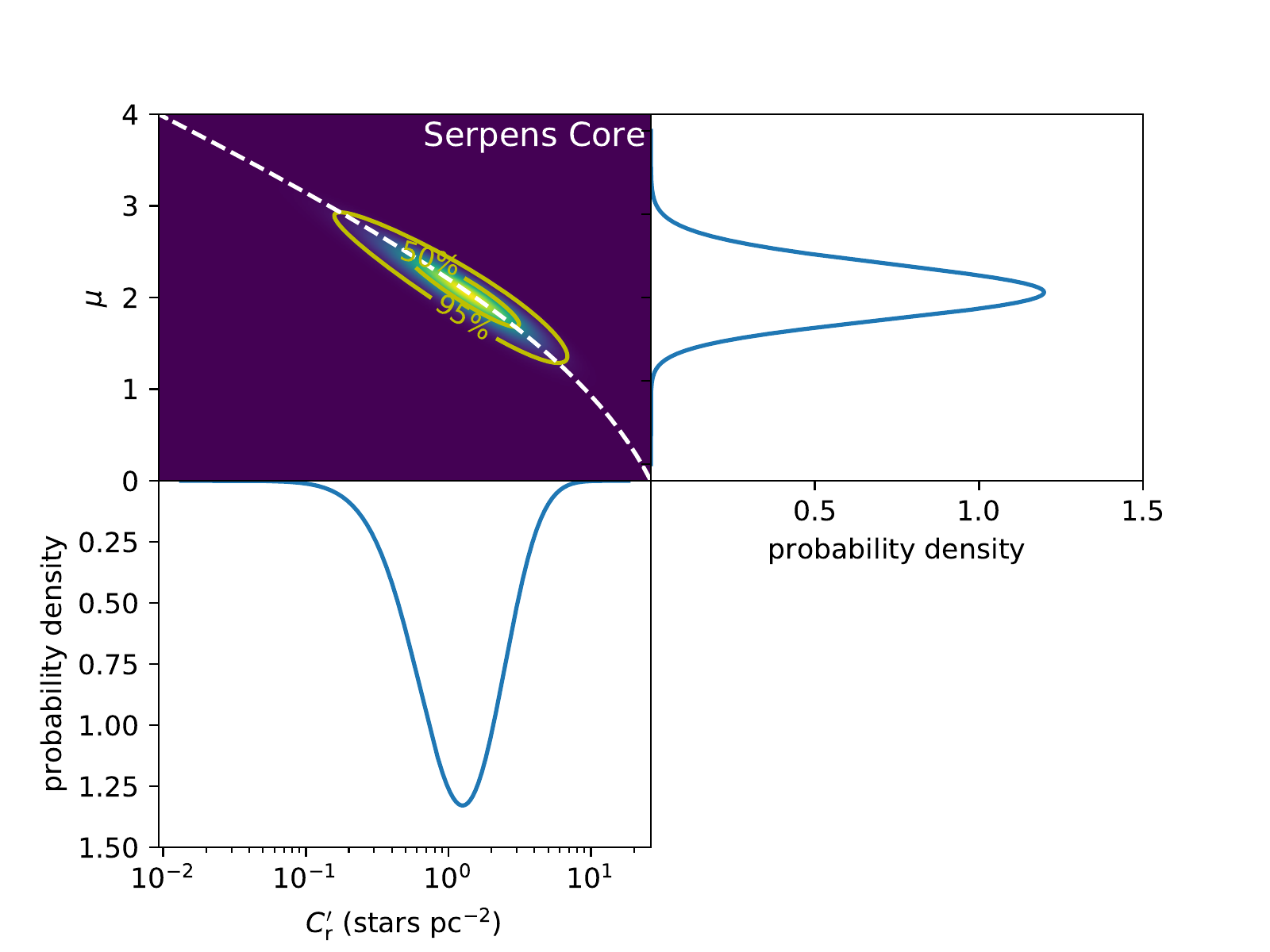}
	\end{subfigure}
	\begin{subfigure}[b]{0.45\linewidth}
		\includegraphics[width=1.1\linewidth]{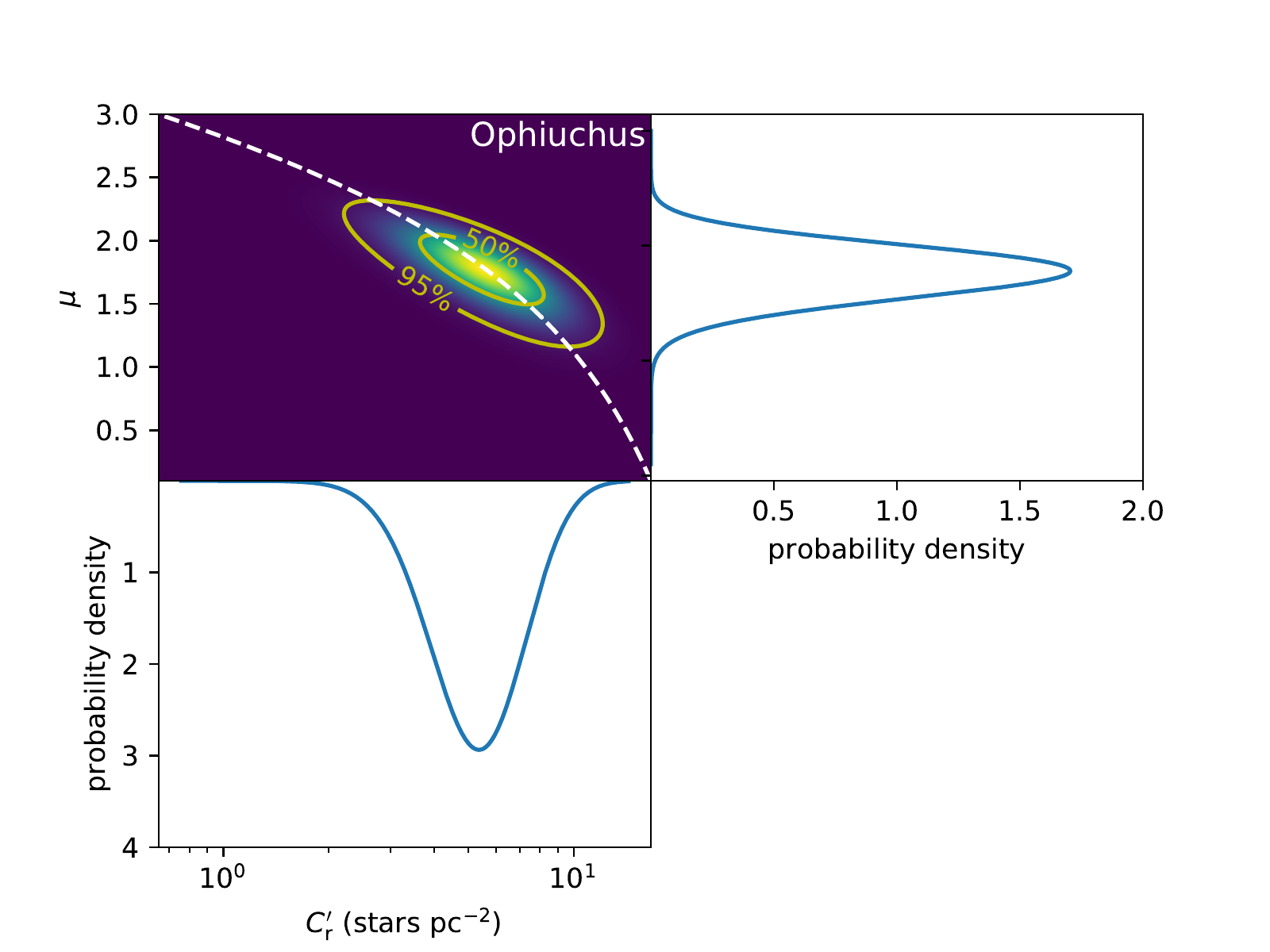}
	\end{subfigure}
	\begin{subfigure}[b]{0.45\linewidth}
		\includegraphics[width=1.1\linewidth]{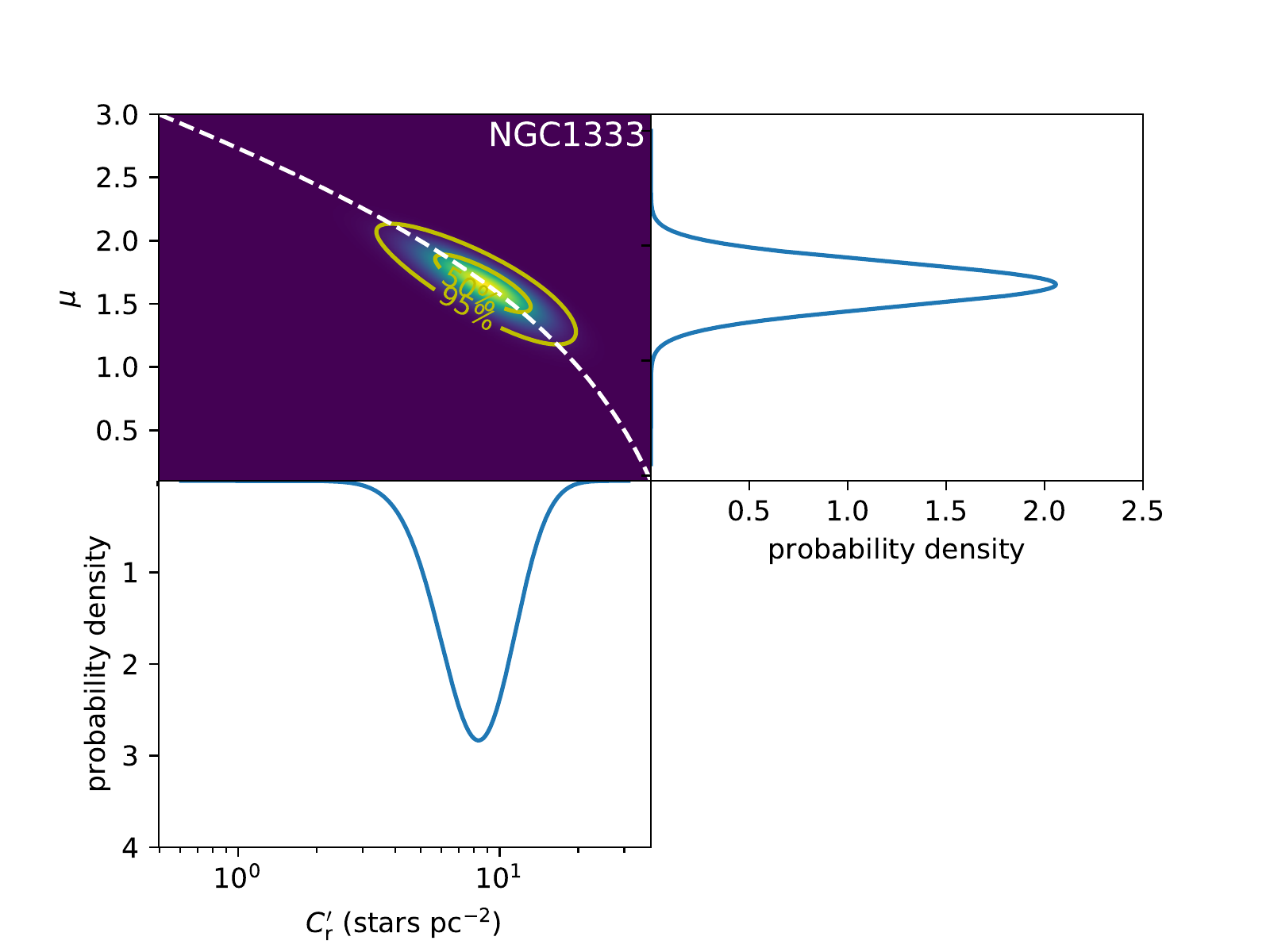}
	\end{subfigure}
	\begin{subfigure}[b]{0.45\linewidth}
		\includegraphics[width=1.1\linewidth]{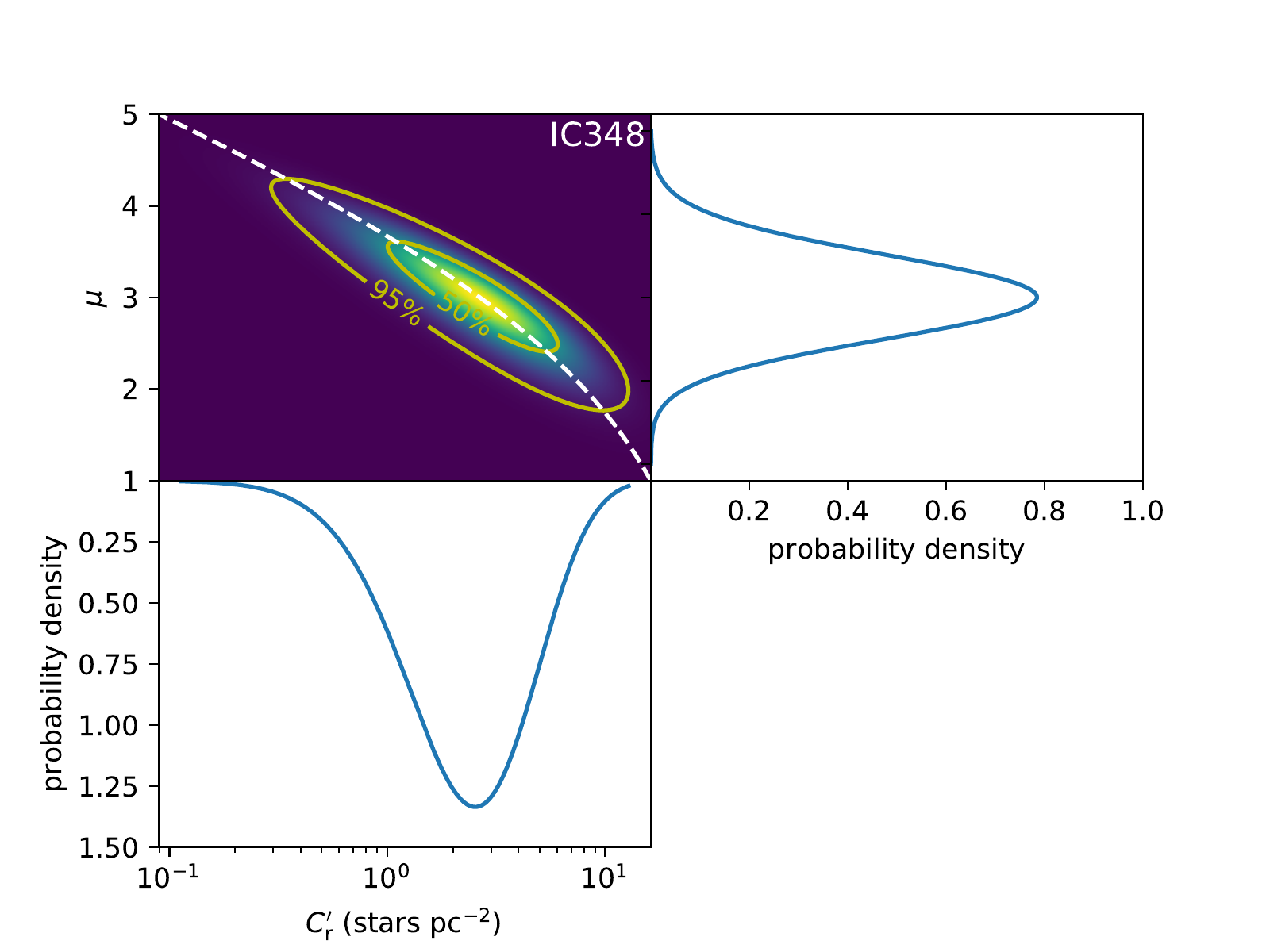}
	\end{subfigure}
\caption[ Joint-probability distribution of Eqn. \ref{eqn:c,b_one_region} for Class~0/I YSOs in Serpens South, Serpens Core, Ophiuchus, NGC1333 and IC348 with the marginalised probability density functions for $\mu$ and $\Cr^{\prime}$]{\label{fig:join-pdf-all} Joint-probability distribution of Eqn. \ref{eqn:c,b_one_region} for Class~0/I YSOs in the labelled star-forming regions with the marginalised probability density functions for $\mu$ and $\Cr^{\prime}$. The contours outline the 50 and 95 per cent cumulative probabilities and the dashed line follows the solutions to Eqn. \ref{eqn:pois_nbar}. The x-axis is $\Cr ^\prime$ (Eqn. \ref{eqn:cr_prime}), because the joint-probability distributions have been calculated using column density values scaled by a factor of $10^{-22}$. This reduces the span of $\Cr$ values and allows the structure in the joint-probability distribution to be distinguishable. See Section \ref{subsec:cr_estimate}}
\end{figure*}}

\begin{table}
\caption{\label{table:mu} $\mu$ estimates for all regions.}
\begin{tabular}{c c c}
Region & Best estimate of $\mu$ & 95 per cent confidence interval\\
\hline
Serpens South & 2.39 & $2.06 \geq \mu \geq  2.70$\\
Serpens Core & 2.06 & $1.45 \geq \mu \geq 2.76$\\
Ophiuchus & 1.78 & $1.28 \geq \mu \geq 2.20$\\
NGC1333	 & 1.66 & $1.28 \geq \mu \geq 2.04$\\
IC348 & 3.00 & $2.03 \geq \mu \geq 4.04$\\
All regions & 2.05 & $1.85 \geq \mu \geq 2.25$ \\ 
\hline
\end{tabular}
\end{table}

\subsection{Estimations of $\Cr$}
\label{subsec:cr_estimate}

In this section we present the results of estimating the region-specific constants for each of the five regions.
The best-estimates for $\Cr$ were produced by marginalising the joint-probability distributions over $\mu$, as demonstrated in Fig. \ref{fig:join-pdf-all}.
These results are presented in Table \ref{table:crprime} as $\Cr^\prime$ values, which are related to $\Cr$ by
\begin{equation}
\label{eqn:cr_prime}
\Cr^\prime = \Cr \times (10^{22})^\mu
\end{equation} 
or, equivalently,
\begin{equation}
\hat{\lambda}\mathrm{(\Nh)} = \Cr^\prime \times \left(\frac{\Nh}{10^{22}~\mathrm{cm^{-2}}}\right)^\mu.
\label{eqn:lmda-crprime-model}
\end{equation}
The results are presented in this form because $\Cr$ values scale with $\Nh^{-\mu}$, and with column density values of order $\sim 10^{22}~\mathrm{cm^{-2}}$ the uncertainties in $\mu$ lead to a large range in the magnitudes of potential $\Cr$ values.
It is because of such large ranges of potential $\Cr$ values that the joint-probability densities of Fig. \ref{fig:join-pdf-all} were presented using $\Cr^\prime$.

While $\Cr$ (and $\Cr^\prime$) values are important for estimating the amount of star formation within a region with Eqn. \ref{eqn:lmda-model}, $\Cr$ does not affect where in a cloud the stars will be positioned in the model -- as can be seen by substituting Eqn. \ref{eqn:lmda-model} into Eqn. \ref{eqn:prob_u} -- and the observed number of YSOs in a region can be used when simulating a spatial point pattern.
In addition, care must be taken in the interpretation of the meaning of the values.
If we take the logarithm of Eqn. \ref{eqn:lmda-model},
\begin{equation}
\log(\Sigma_\mathrm{SFR}) = \log(\Cr) + \mu \log(\Sigma_{\mathrm{Gas}}),
\label{eqn:log-lmda}
\end{equation}
we can see that for a given value of $\mu$, $\Cr$ is equal to the expected YSO surface density when the gas density measure is equal to one, in this case $\Sigma_{\mathrm{Gas}} = 1~\mathrm{cm^{-2}}$.
From this we can see that to compare $\Cr$ values is to compare \emph{expected} YSO surface densities at unit $\Sigma_{\mathrm{Gas}}$ and since $\Sigma_{\mathrm{Gas}}$ can be any density measure, $\Cr$ can be measured at any column density. 

\begin{figure}
\includegraphics[width=\linewidth]{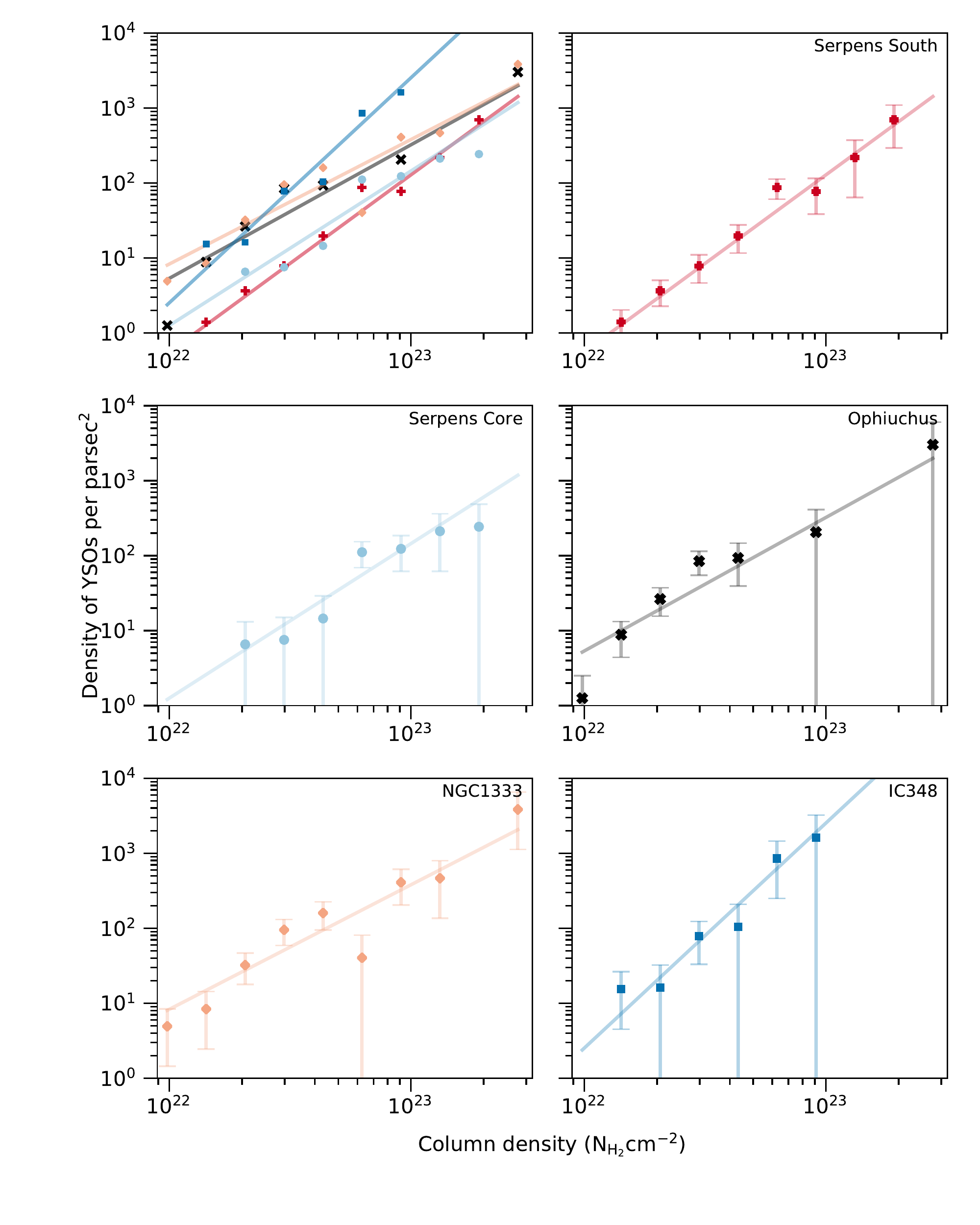}
\caption[YSO surface density measurements within column density bins in Serpens South, Serpens Core, Ophiuchus, NGC1333 and IC348 with straight lines showing best estimates of $\mu$ and $\Cr$ in each region]{\label{fig:yso-best-line} YSO surface density measurements within column density bins in Serpens South, Serpens Core, Ophiuchus, NGC1333 and IC348 with straight lines showing best estimates of $\mu$ and $\Cr$ in each region. 
Uncertainties on Ophiuchus are the Poisson uncertainty on YSO counts to give an idea of YSO surface density uncertainties}
\end{figure}

Fig. \ref{fig:yso-best-line} shows the expected YSO surface densities using the individual best estimates of $\mu$ and $\Cr$ from Tables \ref{table:mu} and \ref{table:crprime} -- this is a reasonable procedure as, in these regions, the best individual estimates of $\mu$ and $\Cr$ are approximately equal to the best joint estimates for $\mu$ and $\Cr$.
From this figure it can be seen that the YSO surface density in these regions is well represented by a power-law with column density using the results of Eqn. \ref{eqn:c,b_all_region}.
As discussed, since $\Cr$ is the expected YSO surface density at a chosen reference column density, we can also see in Fig. \ref{fig:yso-best-line} how the region with the highest $\Cr$ depends on this choice of reference.

\begin{figure}
\includegraphics[width=\linewidth]{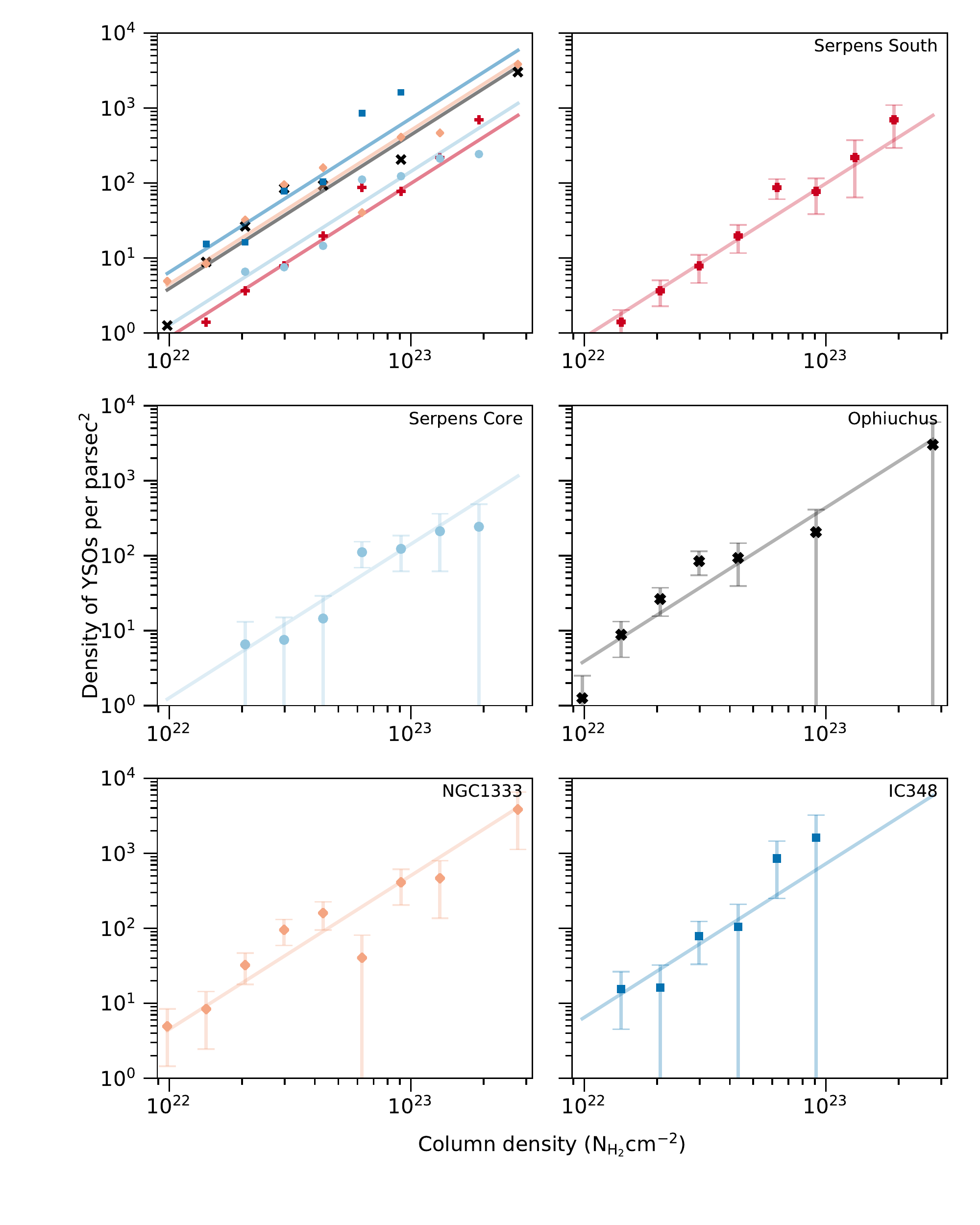}
\caption[ YSO surface density measurements within column density bins in Serpens South, Serpens Core, Ophiuchus, NGC1333 and IC348 with straight lines showing best estimates of $\Cr$ assuming $\mu = 2.05$ in each region]{\label{fig:yso-trend205} YSO surface density measurements within column density bins in Serpens South, Serpens Core, Ophiuchus, NGC1333 and IC348 with straight lines showing best estimates of $\Cr$ assuming $\mu = 2.05$ in each region. 
Error bars are the Poisson uncertainty on YSO counts to give an idea of YSO surface density uncertainties}
\end{figure}

While it is not possible to remove the $\mu$ dependence from $\Cr$, it is possible to find the best-estimates for $\Cr$ in each region assuming the same value of $\mu$ across all regions.  
Table \ref{table:crprime205} presents the best estimates of $\Cr^\prime$ assuming $\mu = 2.05$, and Fig. \ref{fig:yso-trend205} shows the new expected YSO surface density functions.
We can see from these results the effect of different star-formation efficiencies on regions which are assumed to have the same column density dependence. 

\begin{table}
\caption{\label{table:crprime} Estimates of $\Cr^\prime$ from marginalisation over $\mu$ for all regions.}
\begin{center}
\begin{tabular}{c c c}
Region & Best estimate of $\Cr^\prime$ & 95 per cent confidence interval\\
&$(\mathrm{stars~pc^{-2}})$&\\
\hline
Serpens South 	& $0.56$ & $0.31 \geq \Cr^\prime \geq  0.94$\\
Serpens Core 	& $1.26$ & $0.25 \geq \Cr^\prime \geq  3.93$\\
Ophiuchus 		& $5.36$ & $2.71 \geq \Cr^\prime \geq  9.31$\\
NGC1333	 		& $8.31$ & $4.10 \geq \Cr^\prime \geq  14.73$\\
IC348 				& $2.53$ & $0.48 \geq \Cr^\prime \geq  7.5$\\
\hline
\end{tabular}
\end{center}
\end{table}

\begin{table}
\caption{\label{table:crprime205} Estimates of $\Cr^\prime$ for all regions for $\mu = 2.05$.}
\begin{center}
\begin{tabular}{c c c}
Region & Best estimate of $\Cr^\prime$ & 95 per cent confidence interval\\
&$(\mathrm{stars~pc^{-2}})$&\\
\hline
Serpens South 	& $0.88$ & $0.64 \geq \Cr^\prime \geq  1.16$\\
Serpens Core 	& $1.27$ & $0.74 \geq \Cr^\prime \geq  1.93$\\
Ophiuchus 		& $3.91$ & $2.50 \geq \Cr^\prime \geq  5.62$\\
NGC1333	 		& $4.50$ & $3.06 \geq \Cr^\prime \geq  6.22$\\
IC348 				& $6.50$ & $3.11 \geq \Cr^\prime \geq  10.85$\\
\hline
\end{tabular}
\end{center}
\end{table}

\subsection{Application to simulated protostar spatial distributions}
\label{subsec:application_simulated}
\begin{figure*}
\includegraphics[width=\textwidth]{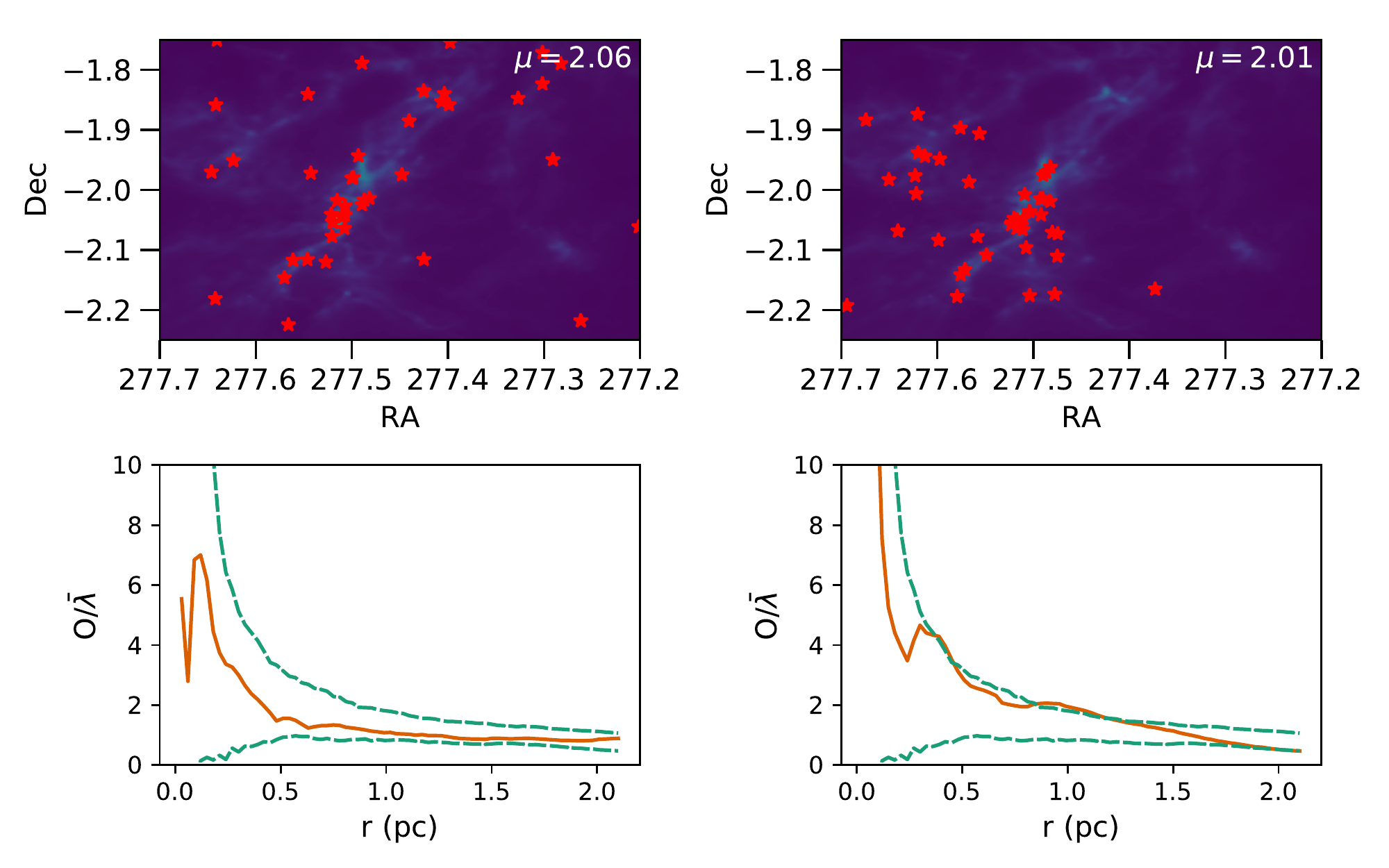}
\caption[Comparison of two distributions of YSOs generated using the \textit{Herschel} column density data for Serpens South and Eqn. \ref{eqn:lmda-model} with $\mu = 2.05$ where one is biased spatially]{\label{fig:serpens_simulated} Two distributions of YSOs generated using the \textit{Herschel} column density data for Serpens South and Eqn. \ref{eqn:lmda-model} with $\mu = 2.05$ with the 95 per cent confidence envelope tests for Eqn. \ref{eqn:lmda-model} with $\mu = 2.05$. The left-hand distribution was generated using the entire study region, while the right, was generated using a probability map covering only the south-west portion of the map.}
\end{figure*}

In this section we apply the O-ring statistic with 95 per cent global confidence envelopes to two sets of simulated YSO distributions in Serpens South, presented in Fig. \ref{fig:serpens_simulated}.
Both sets of simulated data contain the same number of YSO positions as Class~0/I YSOs observed in Serpens South and were generated using Eqn. \ref{eqn:lmda-model} with $\mu = 2.05$ using the \textit{Herschel} column density data for Serpens South.
The left-hand, or unbiased, distribution in Fig.~\ref{fig:serpens_simulated} was generated using a probability distribution which spanned the entire study region, while the right, or biased, distribution was generated using a probability map covering only the south-west portion of the map. 

Using the same methods applied to the star forming regions in Section \ref{subsec:mu}, $\mu$ was measured for the unbiased and biased distributions to find $\mu =2.06$ and $\mu = 2.01$ respectively.

The lower portion of Fig. \ref{fig:serpens_simulated} presents the results of using the O-ring statistic to test for Eqn. \ref{eqn:lmda-model} with $\mu = 2.05$.
We can see from these results that the unbiased distribution is not rejected, and so is consistent with this model -- this is to be expected as the unbiased distribution is a realisation of said model. 
We can also see that the biased distribution rejects this model as the O-ring statistic exceeds the envelope.
These results show how the O-ring test is able to reject the spatially biased distribution of YSOs, whereas a power-law measurement like $\mu$ does not have this type of discriminatory power.

\subsection{Application to protostar spatial distributions}
\label{subsec:application}
The distribution of protostars in Perseus, Ophiuchus and Serpens were tested against four distribution models: a minimum threshold of $6\times10^{21}$ \Nh $\mathrm{cm}^{-2}$ for YSOs to be placed but with no other dependence on column density, a power law dependence of $\mu = 1$, a power law dependence of $\mu = 2.05$, and a power-law equal to the Bayesian-estimated power-law for the region.
The distributions of protostars were tested for their consistency with a distribution model using the O-ring statistic as a summary statistic and 95 per cent global confidence envelopes. 

The O-ring statistic for the distribution of Class~0/I YSOs were measured in each region at a set of spatial scales, 
\begin{equation}
r=\{x | x = n \Delta r \text{, }x \leq R\ \text{, } n \geq 1\}, 
\end{equation}
where $\Delta r = 0.03~\mathrm{pc}$ and $R$ is equal to the half the length of the shortest axis (in either RA or Dec) of the study window.
In other words, $r$ values are linearly spaced in intervals of $\Delta r$ from $r = \Delta r$ up to the largest scale that is less than, or equal to, half of the length of the shortest axis.
Following the results from \citet{retter2019}, the widths of the annuli used in the O-ring test are logarithmic and are equal to 
\begin{equation}
w = 0.6 \times \rho,
\end{equation}
where $\rho$ is the set of spatial scales for each region, $r$, converted into degrees.

Confidence envelopes were generated for each of the four models described earlier using 99 realisations of the first-order processes.
Each realisation was produced by sampling the first-order intensity map a number of times equal to the number of YSOs observed in the region. 
The O-ring statistic was measured for each realisation at the spatial scales $r$ with annuli widths $w$.

All of the confidence envelopes, along with the measured O-ring statistic for the Class0/I YSOs, are presented in Fig. \ref{fig:all_envelopes} and discussed in the following subsections.
The y-axis of the subplots in Fig. \ref{fig:all_envelopes} are the measured O-ring statistic divided by the YSO density of the study window encompassing the star forming region i.e. 
\begin{equation}
\bar{\lambda} = \frac{\sum_u \sum_v \mathrm{P}(u,v)}{\sum_u \sum_v \mathrm{A}(u,v)}.
\end{equation} 
As such the y-axis represents how many more times clustered, or less clustered, the YSOs are compared to CSR in the same window. 

\subsubsection{CSR in gas above cutoff value}
\label{subsubsec:results-csr}
The simplest null hypothesis is that there is no correlation between molecular cloud material and YSOs.
CSR is unlikely to be a successful model as (i) protostars and prestellar cores are known to be associated with dense material within molecular clouds \citep{andre2010}, (ii) the measured power laws in Section \ref{subsec:mu} are greater than zero and (iii) it was shown that Serpens South is inconsistent with this model \citep{retter2019}.

A more sensible, simple relationship between column density and YSOs is one in which some material is required for stars to form but the amount of material has no impact on the number of YSOs. Therefore, we test a model in which star formation is equally likely above some threshold column density, specifically that measured in Taurus, Ophiuchus and Perseus \citep{onishi1998,johnstone2004,andre2010}.
The spatial point process used for the confidence envelopes uses a uniform probability for forming stars in any pixel with a visual extinction above $\mathrm{A_v} = 6$ (assumed to be equal to $6\times10^{21}\, \Nh \mathrm{cm}^{-2}$) and zero otherwise. 

The study windows covered in this paper are too limited in size to come to any conclusions on the existence, or value of, a star-formation column density threshold.
A more complete study of star-formation thresholds would require a greater array of thresholds, and study windows covering more low-column density space. However, including the previously-determined threshold provides a more reasonable model for uniform star formation than no threshold.

The envelopes for this model, presented in the first column of Fig. \ref{fig:all_envelopes}, are exceeded by every region except NGC1333.
For most regions this model produced too few pairs of YSOs at small scales, as evidenced by the measured O-ring statistics exceeding the upper bound of the envelope. 
The O-ring statistic for IC348, however, exceeds the lower bound of the confidence envelope at a separation of 0.7~pc, due to too few YSOs at that separation. These results are summarised in Table~\ref{tab:oring_summary}.

\begin{table}
    \centering
    \begin{tabular}{l |c c c c c}
     & \multicolumn{4}{c}{Class~0/I} &\multicolumn{1}{c}{Class~II}\\
        Model & CSR & $\mu =1$ & $\mu=2.05$ & best $\mu$ & $\mu=2.05$\\
        \hline
        Consistent & 1 & 0 & 3 & 4 & 1\\
        Rejected & 4 & 5 & 2 & 1 & 4\\
    \end{tabular}
    \caption{The number of regions that are consistent with or rejected by the O-ring test for each model.}
    \label{tab:oring_summary}
\end{table}

\subsubsection{Envelopes with $\mu = 1$}
\label{subsubsec:results-mu1}
A power law of $\mu = 1$ means that the surface density of Class~0/I YSOs is directly proportional to the column density.
This is a worthwhile test to perform as it is the simplest relationship in which the surface density of YSOs increases with column density.
It is also of interest as within Orion B the distribution of prestellar cores have been observed to follow a linear relationship with column density above a visual extinction threshold of $\mathrm{A_v} \sim 7$ \citep{konyves2020}, and although the prestellar core distribution is closer to $\mu=2$ in Monoceros~R2 \citep{Sokol2019}, it is still less steep than the protostellar distribution ($\mu = 2.67$).

The results of applying this model are presented in the second column of Fig. \ref{fig:all_envelopes}.
Serpens South, Serpens core, Ophiuchus and IC348 all exceed the envelope at small scales due to YSOs being more clustered at that scale than typically measured with a $\mu$ equal to 1.
NGC1333 also exceeds the envelope though at a more intermediate scale of 1.3~pc. 

\subsubsection{Envelopes with $\mu = 2.05$}
\label{subsubsec:results-mu2}
Following the results discussed previously in Section \ref{subsec:mu} the third model tested was that of the global value of $\mu = 2.05$.
This power is the best estimate of a model where the distribution of Class~0/I protostars is proportional to column density raised to a power which is consistent across the five star-forming regions examined in this paper. 

\begin{figure*}
\begin{center}
\includegraphics[width=\linewidth]{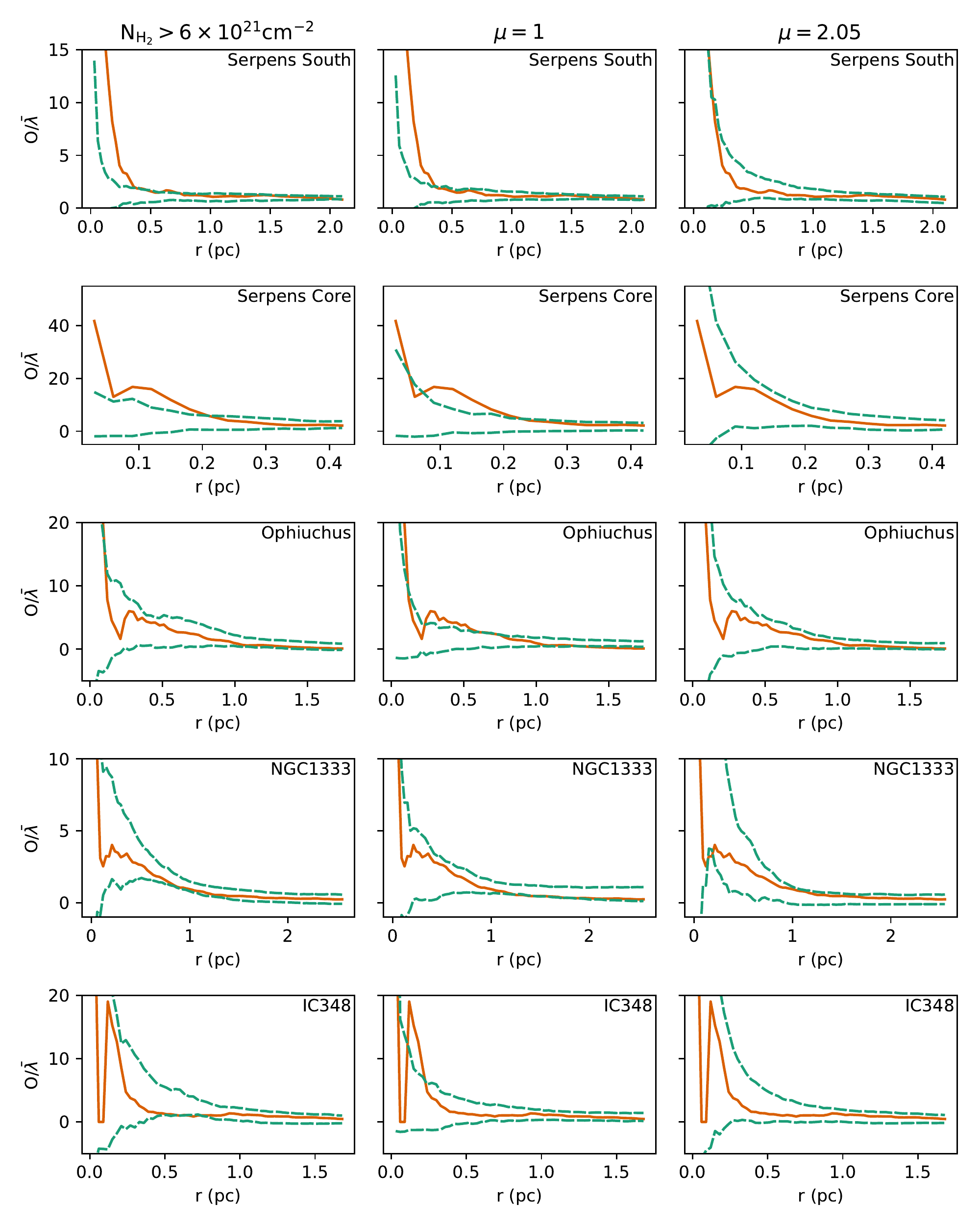}
\caption{\label{fig:all_envelopes} Measured $\mathrm{O}/\hat{\lambda}$ vs $r$ for Class~0/I YSOs in Serpens South, Serpens Core, Ophiuchus, NGC1333 and IC348 with 95 per cent confidence envelopes for different YSO surface-density models: (left) no dependence above a column density threshold of $N_{\mathrm{H}_2} = 6 \times 10^{21} \hbox{cm}^{-2}$; (centre) a power law dependence with $\mu=1$; (right) $\mu=2.05$. }
\end{center}
\end{figure*}

The 95 per cent confidence envelopes presented in the third column of Fig. \ref{fig:all_envelopes} show that Serpens South and NGC1333 both exceed the envelopes at spatial scales around 0.15~pc and therefore reject the model.
IC348, Serpens Core and Ophiuchus remain entirely within the envelopes and are therefore consistent with the model.
While still rejected by two regions, this was the most successful value of $\mu$ tested. 

\subsubsection{Envelopes with best estimate for $\mu$}
\label{subsubsec:results-mubest}
The final test performed on each region was using the best-estimate for $\mu$ calculated in Section \ref{subsec:mu}.
Unlike the previous models where one value of $\mu$ was applied to all of the regions, with this test each region was tested against a different value for $\mu$.

\begin{figure}
\begin{center}
\includegraphics[width=\linewidth]{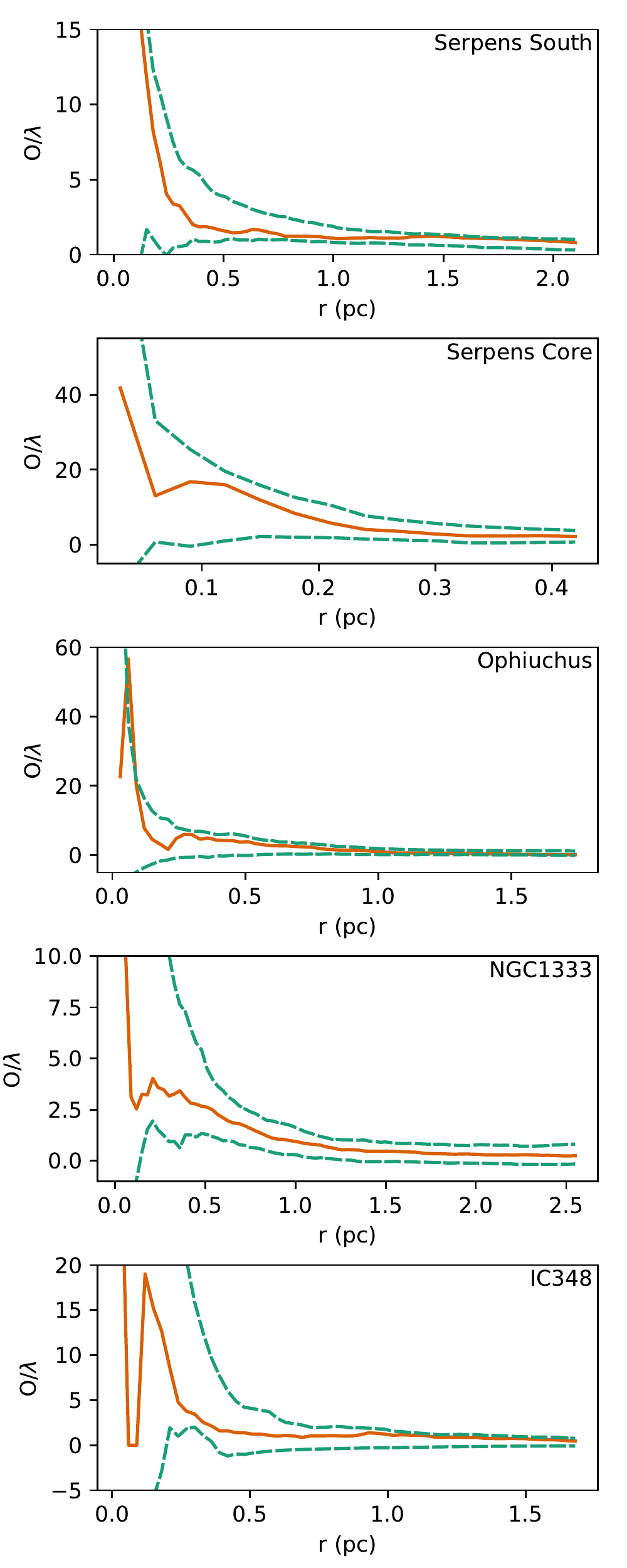}
\caption[Measured $\mathrm{O}/\hat{\lambda}$ vs $r$ for Class~0/I YSOs in Serpens South, Serpens Core, Ophiuchus, NGC1333 and IC348 with 95 per cent confidence envelopes using the best-estimate for $\mu$ in each region from Table \ref{table:mu}]{\label{fig:class0I_bestmu} Measured $\mathrm{O}/\hat{\lambda}$ vs $r$ for Class~0/I YSOs in Serpens South, Serpens Core, Ophiuchus, NGC1333 and IC348 with 95 per cent confidence envelopes using the best-estimate for $\mu$ in each region from Table \ref{table:mu}.}
\end{center}
\end{figure}

Confidence envelopes were produced for each region using the best-estimates of $\mu$ presented in Table \ref{table:mu} and the results are presented in Fig. \ref{fig:class0I_bestmu}.
Serpens South, Serpens Core, NGC1333 and IC348 all remain within their respective envelopes, however Ophiuchus rejects the model on small scales of 0.06~pc.

 \subsection{Application to Class~II YSOs}
\label{subsec:results-classII}
Class~II YSOs are more evolved than Class~0/I sources and tend to be less associated with the dense gas material \citep{mairs2016}; it is likely, then, that the surface-density of Class~II YSOs should follow a different power law with column density to Class~0/I YSOs, if any at all. 
To show that the O-ring statistic with 95 per cent confidence envelopes has enough discriminatory power to distinguish between YSO surface-density models the $\mu = 2.05$ model was applied to the Class~II YSOs in each region. 

The Class~II YSOs were selected from the \citet{dunham2015} catalogue with $-1.6 \leq \alpha	< -0.3$ and $T_\text{bol} > 100\mathrm{K}$.
Due to there being different numbers of Class~II YSOs compared to Class~0/I the confidence envelopes were recalculated using 99 realisations. 

\begin{figure}
\includegraphics[width=\columnwidth]{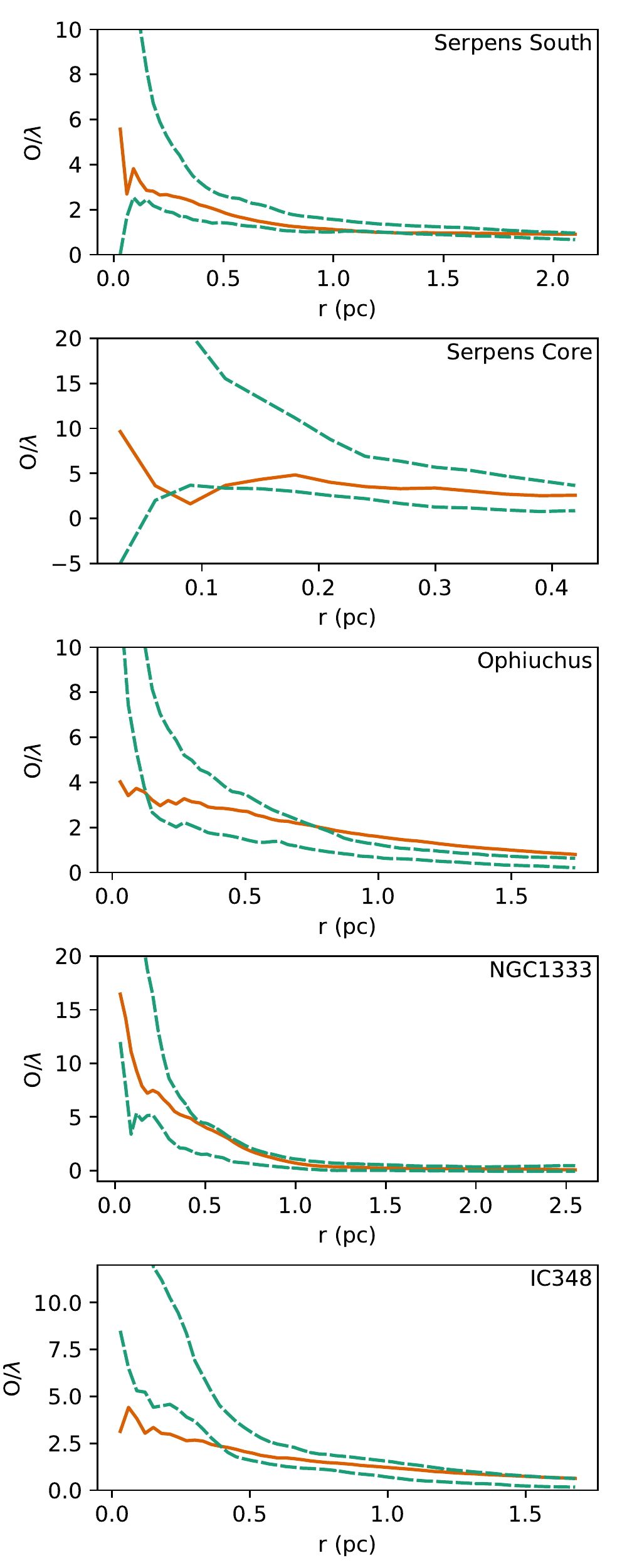}
\caption{\label{fig:classII_envelopes} Measured $\mathrm{O}/\hat{\lambda}$ vs $r$ for Class~II YSOs in Serpens South, Serpens Core, Ophiuchus, NGC1333 and IC348 with 95 per cent confidence envelopes for a $\mu = 2.05$ YSO surface-density model. }
\end{figure}

Presented in Fig. \ref{fig:classII_envelopes} are the measured O-ring statistics and $\mu = 2.05$ model confidence envelopes for the Class~II YSOs in each region. 
The measured $O/\bar{\lambda}$ values show that, except for NGC1333, Class~II YSOs are less clustered (compared to CSR) at small scales than Class~0/I YSOs within the same region.
Serpens South, Serpens Core, Ophiuchus and IC348 exceed the 95 per cent confidence envelopes and therefore reject the $\mu = 2.05$ model, while NGC1333 stays within the envelopes.

\section{DISCUSSION}
\label{sec:discussion}
In Sections \ref{subsec:mu} and \ref{subsec:cr_estimate} parameters relating to possible power law relationships between the column density and the surface density of Class~0/I YSOs were estimated. 
In Section \ref{subsec:application} methods from spatial statistics were used to determine if, and how many, star forming regions from the set were consistent with the stellar distribution models tested.
These are complementary and independent methods as one does not necessarily imply the other; the first test assumed a model and found the parameters that best fit the model, while the spatial statistics tests determined the suitability of the proposed models. 

\subsection{Measured YSO surface density relations}
\label{subsec:disc-measured-mu}

The power-law relationship between the surface density of Class~0/I YSOs and column density was measured in Serpens South, Serpens Core, Ophiuchus, NGC1333 and IC348, the results of which are presented in Tables \ref{table:mu} and \ref{table:crprime}.
The power, $\mu$, was estimated by marginalising the joint-probability distributions of Eqn. \ref{eqn:c,b_one_region} for each region over the region-specific constant, $\Cr$ and vice-versa.
As discussed, $\mu$ defines the relative likelihoods of forming YSOs at different column densities within a region, and $\Cr$ is a region-specific constant which normalises the number of YSOs formed within the region.

The region-specific constants were measured for the individual regions, as discussed in Section \ref{subsec:cr_estimate}; however, it was also discussed that because the region-specific constants depend on $\mu$ and the units of $\Sigma_{\mathrm{Gas}}$ (here,  $N_{\mathrm{H}_2}$), comparison of $\Cr$ between regions of different $\mu$ is difficult to interpret.
The dependence on $\mu$ can be mitigated by considering $\Cr$ values when regions are assumed to have the same $\mu$, and it was shown in Fig. \ref{fig:yso-trend205} that different regions which have the same value of $\mu$ can have different values of $\Cr$.
Such differences in $\Cr$ are due to factors independent of column density that cause different amounts of star-formation, in particular the free-fall time \citep{Pokhrel:2021}.
And so, while $\Cr$ is important for estimating the YSO densities using Eqn. \ref{eqn:lmda-model}, the column density dependence is what is being tested using the O-ring statistic and so discussion will be focused on $\mu$. 

Values of $\mu$ measured for the star forming regions in this work are consistent with studies looking for YSO surface density relationships in other star-forming regions \citep{gutermuth2011, rapson2014,willis2015,lada2013,lombardi2013,lombardi2014,Pokhrel2020}. 
Even high values of $\mu$, such as that of IC348, have been measured such as Perseus with $\mu = 3$ \citep{hatchell2005} and $\mu = 3.8$ \citep{gutermuth2011} and the California Nebula with $\mu = 3.31$ \citep{lada2017} -- though it is shown in Fig. \ref{fig:all_envelopes} and Table \ref{table:mu} that IC348 is consistent with a much lower value of $\mu$.
There is some overlap between the regions tested in this paper and those tested in other works: Ophiuchus with a $\mu$ of 1.78 is exceptionally close to the value of 1.87 and 1.9 measured by \citet{gutermuth2011} and \citet{Pokhrel2020} respectively, and IC348, within the Perseus molecular cloud shows a similar power law to \citet{gutermuth2011}, though NGC1333 does not. Some of the variation in $\mu$ may be an indication of deviations from a $\mu=2.0$ power law at higher column densities \citep{Pokhrel:2021}. 
It is interesting that these power laws show such similarity given the differences in the methods of measuring the power law, the column density measures, identifying the YSOs and the star-forming regions used.
There is even potential that some of the higher values of $\mu$, such as that of IC348, may be reduced in future with increasing resolution as happened with Orion B \citep{lombardi2014}.

In addition to measuring $\mu$ for each region, Eqn. \ref{eqn:c,b_all_region} was used to estimate the power-law value which best represents the YSO distributions in all five regions simultaneously by marginalising over the region-specific constants: $\mu = 2.05$.
Unlike taking an average value of $\mu$, which requires measured values of $\mu$ and an assumption as to how they should be weighted, this method directly uses the available data to estimate the parameter. 
Given this difference in methodology, it is interesting how similar this value is to the weighted mean $\mu = 2.06 \pm 0.14$ (with 95 per cent confidence intervals) for these regions and the mean value of $\mu = 2.0$ for the 12 regions studied in \citet{Pokhrel2020, Pokhrel:2021}.

Examining the individual estimates of $\mu$ presented in Table \ref{table:mu} and Fig. \ref{fig:mu_regions}, $\mu = 2.05$ appears to represent the ensemble of Bayesian fitted $\mu$ values well. 
While this is expected, given that this value of $\mu$ was estimated using the YSO and area counts that were used to produce the values of $\mu$ for each region, it did not use the values of $\mu$ themselves and so demonstrates that the combination of measurements produces a value that is reasonable and representative.
We can also see in Fig. \ref{fig:yso-trend205} that for most regions $\mu = 2.05$, combined with the appropriate estimate of $\Cr$, provides a good, visual representation of the YSO surface density measurements despite not being the best-estimate for the most regions.

\subsection{Testing YSO distributions against spatial distribution models}

A measured $\mu$ value describes how the surface density of YSOs changes in general across the entire study region.
Measuring a power-law, however, does not mean that the YSOs are evenly distributed according to column density throughout the cloud.
This was demonstrated in Section \ref{subsec:application_simulated}, where a value of $\mu$ measured from an evenly distributed population of YSOs could be reproduced in a population of stars distributed over only half of the cloud. Hence it is reasonable to say that $\mu$ is a useful metric to describe YSO distributions but is not enough on its own to say whether YSOs have a relationship with cloud material of the form Eqn. \ref{eqn:lmda-model}.
By utilising the spatial information, spatial statistics can test if observed distributions of YSOs are consistent with a power-law relationship with column density. 

\subsubsection{Class~0/I YSOs}

One should expect the surface density of YSOs to be affected by column density. 
From a physics standpoint this makes sense as a greater reservoir of material has the potential to form more stars, and from an observational standpoint the values of $\mu$ measured for star-forming regions are all greater than zero.
The O-ring tests confirm this as Serpens South, Serpens Core, Ophiuchus and IC348 all have Class~0/I YSO populations inconsistent with YSOs positioned independently of column density above $\mathrm{A_v} = 6$.
It was also confirmed by the O-ring test that the relationship between $\mu$ and column density is likely superlinear as all five regions rejected $\mu = 1$ and subsequent tests with higher values of $\mu$ all had fewer rejections (see Table~\ref{tab:oring_summary}).

Each region was also tested against the global estimate of $\mu = 2.05$, the 95 per cent confidence envelopes for which are presented in Fig. \ref{fig:all_envelopes}.
These results show that of the five regions tested, Serpens Core, Ophiuchus and IC348 have Class~0/I YSO populations that are consistent with the $\mu = 2.05$ model.
While it is unsurprising that Serpens Core is consistent with $\mu = 2.05$, given its power law was estimated to be $\mu = 2.06$, this is a more interesting result for Ophiuchus and IC348 as their estimates for $\mu$ were $1. 78$ and $3.00$ respectively. 

O-ring tests for Serpens South and NGC1333 rejected the $\mu = 2.05$ model.
This was due to overclustering and regularity for Serpens South and NGC1333 respectively. 
Interestingly, the outcome of the envelope tests -- with Serpens South and NGC1333 rejecting the $\mu = 2.05$ model while the other regions do not -- is mirrored in the $\mu$ values measured in Table \ref{table:mu}.
The power $\mu = 2.05$ is within the 95 per cent confidence intervals for Serpens Core, Ophiuchus and IC348 individually while it is marginally outside the interval for Serpens South and NGC1333. 
It is perhaps due to the proximity of $2.05$ to the 95 percent confidence intervals of Serpens South and NGC1333 that the O-ring values exceed the confidence envelopes over such a small set of spatial scales at $\sim 0.15~\mathrm{pc}$.

Finally, each region was tested against its best-estimate for $\mu$. 
Unlike the other models which assume a single value of $\mu$, this model contains $\mu$ as an adjustable parameter for each region.
By having four additional adjustable parameters in total, one should expect the number of YSO distributions that are consistent with the model to increase.
This was observed in Fig. \ref{fig:class0I_bestmu} where it was found that Serpens South, Serpens Core, NGC1333 and IC348 all have YSO populations consistent with their best-estimates of $\mu$.
From these results then we can see that Eqn. \ref{eqn:lmda-model}, using the Bayesian estimates of $\mu$, is generally supported by spatial statistics.
Though Ophiuchus rejected $\mu = 1.78$ at 0.06~pc, on similar scales to the regions which rejected $\mu = 2.05$.

\subsubsection{Class~II YSOs}
While $\mu$ values were not measured for the Class~II YSOs in these regions, by looking at the measured O-ring statistic and $\mu=2.05$ envelopes in Fig. \ref{fig:classII_envelopes} it is clear that the two populations are not equally dependent on column density. 
The Class~II YSOs in Serpens South, Serpens Core, Ophiuchus and IC348 are all inconsistent with the a $\mu=2.05$ model, while those in NGC1333 remain within the envelope.
This increase in rejection by more evolved YSOs, in combination with lower $O/\bar{\lambda}$ values and generally flatter O-ring results as a function of radial separation, shows that there is a change in the separation of YSOs as a function of their age. 
These results also demonstrate that these tests have enough discriminatory power to distinguish between two distinct but related populations within the same region - Class~0/I and Class~II YSOs.

\subsection{Potential for a universal column density model}
\label{subsec:disc-evidence_global}

A question proposed at the beginning of this work was if it is possible to describe the locations of YSOs within a molecular cloud with a model that only uses column density.
After applying four different models to the Class~0/I YSOs in five star forming regions every region was found to be consistent with at least one model (see Table~\ref{tab:oring_summary}).
The answer to this question, therefore, appears to be `yes' as the parameters for a given model can be tweaked in order to be consistent with a 
given set of YSOs.
Given that an individual region can be described using a column density model, the next question is whether it is possible to describe the distributions of YSOs within multiple molecular clouds using the same column density model. 

The most successful of the four models tested was that in which the best-estimate of $\mu$ calculated for each region using the Bayesian methodology from Section \ref{sec:bayes_stats} was applied.
Using this model, four out of the five regions were found to have YSO distributions consistent with being distributed throughout the cloud according to column density alone. 
It is possible, therefore, that if YSOs are distributed following column density alone that $\mu$ simply varies between star-forming regions and that there is no universal power-law distribution. 
However, not all of the regions were consistent with their best-estimate of $\mu$ and it is difficult to say whether this increase in consistency with the data is significant enough to justify the addition of an adjustable parameter to the model. 

As discussed in Sections \ref{subsubsec:results-mu2} and \ref{subsec:disc-measured-mu}, multiple regions can be consistent with the same power-law despite the best estimate of their $\mu$ values not being equivalent.
Fig. \ref{fig:yso-trend205} shows how a YSO surface density proportional to column density to the power of $\mu = 2.05$ represents the data quite well, and using the O-ring statistic the $\mu=2.05$ model is able to describe the YSO distributions of Serpens Core, Ophiuchus and IC348 across all of the tested spatial scales.
Out of the three models tested using a single value of $\mu$, $\mu = 2.05$ performed the best with three regions out of five being consistent with the distribution.
The first test with $\mu = 0$ above a column density threshold was only consistent with NGC1333 and $\mu = 1$ was not consistent with any of the regions. 

\begin{figure}
\includegraphics[width=\columnwidth]{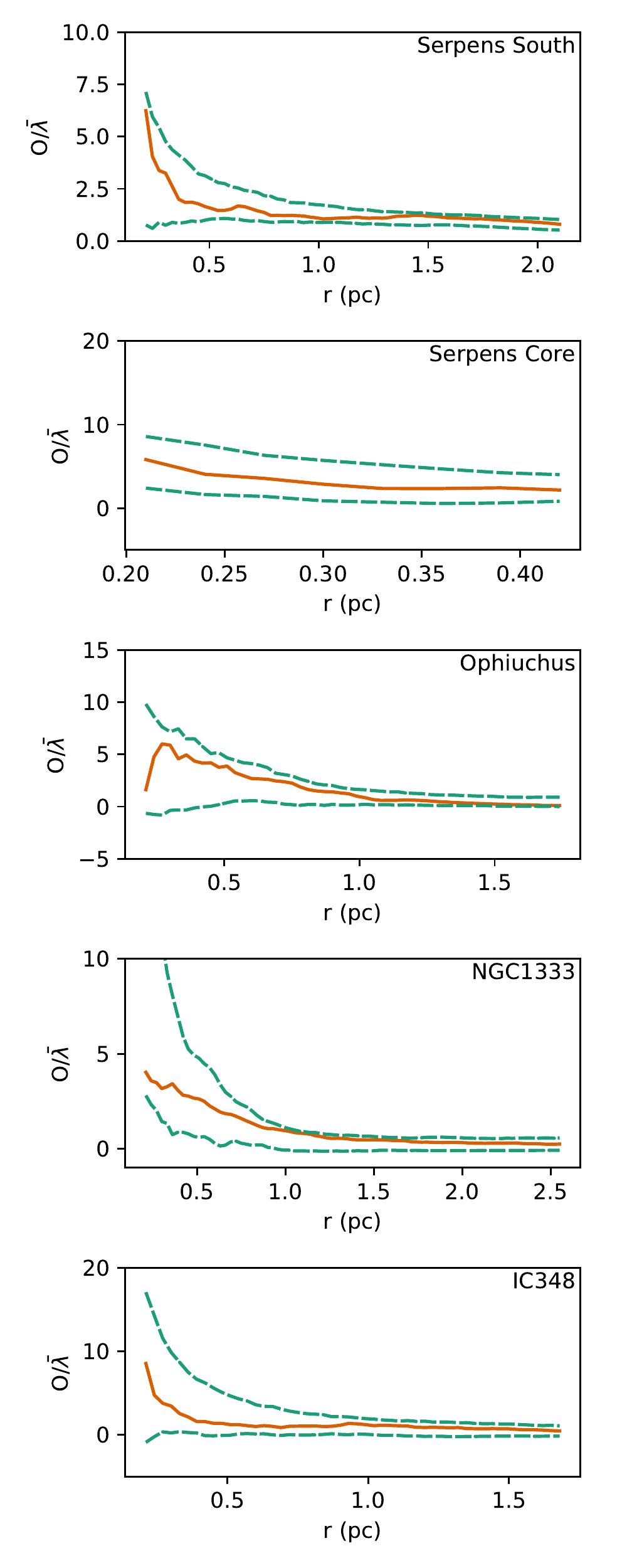}
\caption{\label{fig:mu2_larger} The measured $\mathrm{O}/\hat{\lambda}$ vs $r$ for Class~0/I YSOs in Serpens South, Serpens Core, Ophiuchus, NGC1333 and IC348 from Fig. \ref{fig:all_envelopes} with 95 per cent confidence envelopes for a $\mu = 2.05$ YSO surface density model, using $r > 0.18~\mathrm{pc}$.}
\end{figure}

While the Class~0/I YSOs in Serpens South and NGC1333 rejected the $\mu = 2.05$ model, this rejection was only over a small set of spatial scales between $0.12~\mathrm{pc}$ and $0.18~\mathrm{pc}$, and on other scales the distribution was consistent with the model.
This can be seen in Fig. \ref{fig:mu2_larger} which shows the measured O-ring data from Fig. \ref{fig:all_envelopes} for $r > 0.18~\mathrm{pc}$; the O-ring statistics in Fig. \ref{fig:mu2_larger} remain within the envelope across all scales and so appear consistent with the $\mu = 2.05$ model.
It is worth emphasising that, while the envelopes in Fig. \ref{fig:mu2_larger} have been adjusted compared with the right-hand panels of Fig. \ref{fig:all_envelopes} to retain a 95 per cent significance level, the envelopes have been calculated using the same null hypothesis data and so are not an independent test.
Fig. \ref{fig:mu2_larger} does show, however, that remaining within the envelopes at larger $r$ values is a feature of the data and not due to the envelopes being widened by the O-ring values which exceed the envelopes.
From this additional check we can say that the large-scale behaviour of the Class~0/I YSOs in all of these regions is well described by the same power-law relationship with column density.

\subsection{Alternative universal models}

Using both spatial statistics and Bayesian statistics it was shown that a power-law model with $\mu = 2.05$ provides a good approximation to the data.
It is interesting to note that while Eqn. \ref{eqn:lmda-model} appears to fit the measured YSO surface density data in Serpens South and NGC1333, as shown in Fig. \ref{fig:yso-trend205}, these regions both exceed the confidence envelopes. 
This could imply a situation like that discussed in Section \ref{subsec:application_simulated} where column-density-independent effects resulted in star formation being unevenly distributed throughout the cloud.

The surface density of YSOs is not necessarily proportional to the column density to some power and so some modification of the first-order model Eqn. \ref{eqn:lmda-model} may produce better results.
The number of different models which could be simulated are potentially unlimited; however, the excursions from the $\mu = 2.05$ envelope were brief and the data elsewhere are consistent with the power law and so any additional changes to the model should not have a large effect on the power-law relationship.
Furthermore the scales on which these additional parameters influence the star formation need only be limited to small scales. 

The results in Fig. \ref{fig:all_envelopes} show that Serpens South and NGC1333 reject the model at scales close to 0.15~pc due to the over-clustering and under-clustering respectively and are otherwise consistent with the model.
At most scales, therefore, the distribution of YSOs in Serpens South and NGC1333 behave similarly to this simple power-law relationship -- except at a spatial scale of 0.15~pc.
From Fig. \ref{fig:all_envelopes} is not possible to determine exactly what this rejection means without further testing, however some possibilities will be discussed here.

The first option is that a global first-order model for star formation between clouds requires a different power law.
This appears unlikely as the O-ring statistic shows over-clustering in Serpens South and under-clustering in NGC1333.
Any increase in $\mu$ would lead to increased clustering at smaller scales while a decrease in $\mu$ would have the opposite effect, neither of which would necessarily represent Serpens South and NGC1333 simultaneously. 
Fig. \ref{fig:all_envelopes} shows the results for a power law of $\mu = 1$ which consistently under-represent the density in Serpens South while NGC1333's O-ring statistic is consistent at small scales.
Extrapolating the envelopes between the first and second columns of Fig. \ref{fig:all_envelopes} provides some evidence against power laws less than 2, though more simulations would be required to conclusively determine this to be true.

A second option would be to add more parameters to the surface density model. 
One parameter of particular interest is a column density threshold for star formation.
Indeed, from Fig. \ref{fig:all_envelopes} it was shown that the YSOs within NGC1333 are consistent with being positioned randomly in pixels with a column density greater than $6 \times 10^{21} \mathrm{cm}^{-2}$.
\citet{lombardi2013}, hereby LLA, introduced a Bayesian method related to that in this paper which uses the positions of protostars and the visual extinctions at those positions to estimate parameters in their model for protostellar surface densities.
The surface density model in LLA is similar to Eqn. \ref{eqn:lmda-model}, except with two parameters in addition to $\Cr$ and $\mu$ (in their notation $\kappa$ and $\beta$ respectively): $\sigma$ and $A_0$.
$\sigma$ is a diffusion coefficient term which allows for some amount of travel between the protostars' sites of formation and observation and $A_0$ is a star formation threshold density.
\citet{lada2013} applied the method of LLA to Orion A, Orion B, California and Taurus, and found that there was no significant measurement of a diffusion coefficient and that a star-formation threshold may be more due to the distribution of material in the cloud -- suggesting that the model is scale free.
From the results in this paper it is not possible to come to a conclusion on a model which uses both a power law and a threshold; no such model was tested and the number of YSOs from low column density regions in this work is insufficient to provide much insight on YSO distributions at low column densities.
However, given the results of LLA the effect of including a column density threshold would be likely to be limited. 

A third option is that the power-law model cannot be applied to small spatial scales.
This could be due to data-related problems, for example resolution. 
The spatial separations used start at, and are separated by, an interval of 0.03~pc, at these small spatial scales resolution effects become more prominent which in turn increases the likelihood that close, separate sources will be counted as a single source or vice-versa.  This is a particular problem at small radii where a small change in the number of YSOs has a large impact on the density.  
It could also be that the distributions of YSOs are affected by different physics at small-scales than large scales.
The scales at which Serpens South and NGC1333 reject the global power law are at scales close to the filament scale of $0.1$~pc \citep{arzoumanian2011}, and average core separations in filaments of 0.14~pc \citep{konyves2020}.
Given that YSOs form within collapsing filaments it is possible that the structure of the filaments in which these Class~0/I YSOs form affects their distributions.  

Finally, it may also be the case that a model with second-order components is needed to capture the nature of the distribution of star formation in star-forming regions.
In a first-order model the clustering of YSOs is a product of a general increase in YSO density due to a change in environment, in this case column density.
It is also possible that clusters of protostars are not simply a function of increased density but are instead a product of a cluster-formation process which preferentially generates clusters in higher-density regions.
Such behaviour can be represented through application of second-order effects which raise or lower the probability of forming a star as a function of distance from another star.
This could be a YSO disrupting the column density of its immediate surroundings for example in NGC1333 \citep{knee2000}, or it could be clusters of protostars forming within a dense core or filament as, for example, in Perseus \citep{Tobin2016}.  

\subsection{Changing evolutionary timescales with column density}
\label{subsec:disc-mu1}

As discussed in Section \ref{subsubsec:results-mu1}, the surface density of prestellar cores in Orion B has been observed to follow a less steep relationship with column density compared to the YSOs in Orion B, which follow a power-law with $\mu \approx 2$ \citep{lombardi2014,konyves2020,Pokhrel2020}.
Similarly, studies on the prestellar and protostellar populations in Monoceros R2, Serpens South, Ophiuchus and Perseus showed that the protostars in these regions were more clustered than their prestellar counterparts \citep{Enoch2008, gutermuth2011, Sokol2019}.
This leads naturally to the question of why these power laws should be different if prestellar cores are expected to evolve into Class 0 YSOs. 
One reason for this could be that the prestellar cores and Class~0/I YSOs have timescales that are affected by environment in different ways, or, the converse argument, for these distributions to be the same it would require the evolutionary timescales of prestellar cores and Class~0/I YSOs to share the same dependence on the environment.

To see why this is the case a simple model of the rates of change of surface density over time, similar to that in nuclear decay, is introduced using a subset of eqns. (2)--(7) from \citet{kristensen2018}. 
Assuming prestellar cores are produced at a constant rate, $\gamma$ (which may be a function of local column density), and evolve into Class~0/I YSOs with a lifetime $\tau_{\mathrm{PC}}$ the change in prestellar core surface density is
\begin{equation}
\label{eqn:dpc/dt}
\frac{\mathrm{d}\Sigma_{\mathrm{PC}}}{\mathrm{d}t} = \gamma - \frac{\Sigma_{\mathrm{PC}}(t)}{\tau_{\mathrm{PC}}},
\end{equation}
where $\Sigma_\mathrm{PC}$ is the surface density of prestellar cores. 
Similarly, assuming Class~0/I YSOs evolve into Flat or Class~II YSOs with a lifetime $\tau_{\mathrm{0/I}}$ the surface density of Class~0/I YSOs, $\Sigma_{\mathrm{0/I}}$, is
\begin{equation}
\label{eqn:d0i/dt}
\frac{\mathrm{d}\Sigma_{\mathrm{0/I}}}{\mathrm{d}t} = \frac{\Sigma_{\mathrm{PC}}(t)}{\tau_{\mathrm{PC}}} - \frac{\Sigma_{\mathrm{0/I}}(t)}{\tau_{\mathrm{0/I}}}.
\end{equation}
The solutions to Eqns. \ref{eqn:dpc/dt} and \ref{eqn:d0i/dt} are 
\begin{equation}
\label{eqn:sigma_pc}
\Sigma_{\mathrm{PC}} = \gamma\tau_{\mathrm{PC}} \left( 1 - e^{-t/\tau_{\mathrm{PC}}}\right),
\end{equation}
and
\begin{equation}
\label{eqn:sigma_01}
\Sigma_{\mathrm{0/I}} = \gamma\tau_{\mathrm{0/I}} \left( 1 - \frac{\tau_{\mathrm{PC}}}{\tau_{\mathrm{PC}}-\tau_{\mathrm{0/I}}}e^{-t/\tau_{\mathrm{PC}}} - \frac{\tau_{\mathrm{0/I}}}{\tau_{\mathrm{0/I}}-\tau_{\mathrm{PC}}}e^{-t/\tau_{\mathrm{0/I}}}\right)
\end{equation}
respectively, where it is assumed that at $t = 0$ the surface density of prestellar cores and protostars are zero.

Everything inside the brackets of Eqns. \ref{eqn:sigma_pc} and \ref{eqn:sigma_01} is unitless and column density independent.
The column density dependence of a population, therefore, is defined by the product of prestellar core formation rate and the lifetime of the population.
For simplicity the solutions to the steady-state condition, where Eqns. \ref{eqn:dpc/dt} and \ref{eqn:d0i/dt} are both equal to zero, are
\begin{equation}
\label{eqn:pc_steady}
\Sigma_{\mathrm{PC}} = \gamma\tau_{\mathrm{PC}}
\end{equation}
and
\begin{equation}
\Sigma_{\mathrm{0/I}} = \gamma\tau_{\mathrm{0/I}}.
\label{eqn:0I_steady}
\end{equation}
From inspection it can be seen that for $\Sigma_{\mathrm{PC}}$ and $\Sigma_{\mathrm{0/I}}$ to share the same column density dependence, their lifetimes must also be equally dependent on column density. 
In other words, if the prestellar cores and Class~0/I YSOs in a region have different column density dependences this could be due to different column density dependences of the evolutionary timescales of prestellar cores and protostars.

As discussed, observations in Orion B find that $\Sigma_{\mathrm{PC}}$ is linearly proportional to $\Sigma_{\mathrm{Gas}}$ while $\Sigma_{\mathrm{0/I}}$ is proportional to $\Sigma_{\mathrm{Gas}}$ to a power of about two; additionally, measurements in Monoceros R2 show $\Sigma_{\mathrm{PC}} \propto \Sigma_{\mathrm{Gas}}^2$ \citep{Sokol2019, Pokhrel2020,Pokhrel:2021} and $\Sigma_{\mathrm{0/I}} \propto \Sigma_{\mathrm{Gas}}^{2.67}$ \citep{gutermuth2011}.
And so, while the exact values of $\mu$ may differ between regions, the fact that $\mu$ varies between classes, combined with Eqns. \ref{eqn:pc_steady} and \ref{eqn:0I_steady}, strongly suggests that $\tau_{\mathrm{PC}}$ and $\tau_{\mathrm{0/I}}$ must have different dependencies on column density due to interactions with the environment such as ongoing accretion.

To gain some insight into how the relative time-scale depends on column density in Orion B we substitute in the observed relations of $\Sigma_{\mathrm{0/I}} \propto \Sigma_{\mathrm{Gas}}^2$ -- from this and other measurements discussed in Section \ref{subsec:disc-evidence_global} -- and $\Sigma_{\mathrm{PC}} \propto \Sigma_{\mathrm{Gas}}$ from \citet{konyves2020}:
\begin{align}
\label{eqn:gamma_taupc}
\gamma\tau_{\mathrm{PC}} &\propto \Sigma_{\mathrm{Gas}} ,\\
\gamma\tau_{\mathrm{0/I}} &\propto  \Sigma_{\mathrm{Gas}}^2
\end{align}
and
\begin{equation}
\label{eqn:ratio3}
\frac{\tau_{\mathrm{PC}}}{\tau_{\mathrm{0/I}}} \propto \Sigma_{\mathrm{Gas}}^{-1},
\end{equation}
where Eqn. \ref{eqn:ratio3}, the ratio of $\Sigma_{\mathrm{PC}}$ and $\Sigma_{\mathrm{0/I}}$, states that the difference in column density dependence between $\tau_{\mathrm{PC}}$ and $\tau_{\mathrm{0/I}}$ is a factor of $\Sigma_{\mathrm{Gas}}$. 
This suggests that prestellar cores evolve more quickly at higher column densities than Class~0/I YSOs. 
There are different ways to interpret this: (i) prestellar cores evolve on shorter time-scales at higher column densities; (ii) Class~0/I YSOs remain embedded in their envelope longer at higher column densities; (iii) alternatively, both are column density dependent in some form with prestellar cores ultimately evolving faster than Class~0/I YSOs at higher column densities.

It is very likely that both $\tau_{\mathrm{PC}}$ and $\tau_{\mathrm{0/I}}$ are column density dependent.
For prestellar cores their lifetime is often compared to the free-fall time of a spherically-symmetric mass,
\begin{equation}
\label{eqn:free-fall}
\mathrm{t_{ff}} \propto \rho^{-1/2},
\end{equation}
where $\rho$ is the density of the sphere.
Eqn. \ref{eqn:free-fall} shows that, since free-fall time is proportional to volume density to a power $-1/2$, higher density objects collapse more quickly.
Numerical simulations have shown that Bonner--Ebert spheres have higher central densities and are quicker at collapsing within higher density environments \citep{Kaminski2014}. 
Observationally, the smaller numbers of prestellar cores observed with higher densities also suggest that lifetime decreases with increasing density \citep{jessopwt00,konyves2015}.
Finally, normalising the column density by the free-fall time results in a linear relationship between $\Sigma_\mathrm{gas}/t_\mathrm{ff}$ and $\Sigma_\mathrm{SFR}$ \citep{Pokhrel:2021}.
Hence it is likely that $\tau_{\mathrm{PC}}$ is lower at higher column densities.

For Class~0/I protostars to take longer to evolve at higher column densities it would require that they remain embedded within their envelopes for longer compared to their lower-column-density counterparts.
It may be the case that Class~0/I protostars are able to remain embedded while material is available for accretion, which would result in longer lifetimes in regions that are more dense \citep{hatchell08}.
This is in part supported by numerical simulations where it was found that the accretion rate onto protostars was equivalent between two simulated clouds of different densities \citep{2005MNRAS.356.1201B}.
If Class~0/I protostars do take longer to evolve at higher column densities,
a change in $\tau_{\mathrm{0/I}}$ with respect to column density could be observable in the relative masses in protostars in regions of different column density. 
Indeed some evidence of this has been observed in mass segregation in YSOs and dense cores, where the most massive sources were found within regions with higher densities of sources and towards the central location of the cluster \citep{kirk2011,Kirk_2016}. 
It was also noted in \citet{2005MNRAS.356.1201B} that objects formed within a denser cloud showed a greater variation in the time taken for an object to accrete. 
As a counter argument, the same simulations also showed that dynamical interactions between objects were the dominant force in terminating accretion and objects were more likely to be ejected sooner in a higher density cloud \citep{bate2012}.
This would imply that $\tau_{\mathrm{0/I}}$ is smaller in higher column densities.
These are, however, results from numerical simulations and observational evidence is currently insufficient to convincingly support either lengthening or shortening Class~0/I lifetimes.

Ultimately, it is not possible to determine which of the three terms $\gamma$, $\tau_{\mathrm{YSO}}$ or $\tau_{\mathrm{0/I}}$ are column density dependent from Eqns. \ref{eqn:gamma_taupc} -- \ref{eqn:ratio3}.
However, a minimum of two of the terms must be functions of column density, at least one of which must be an evolution time-scale for prestellar cores or Class~0/I YSOs. 
This is true for any region in which $\Sigma_{\mathrm{PC}}$ and $\Sigma_{\mathrm{0/I}}$ are measured to have different dependencies on column density.

\section{CONCLUSIONS}

In this paper the distribution of Class~0/I YSOs in Serpens South, Serpens core, Ophiuchus, NGC1333 and IC348 were tested against a spatial distribution model of the form
\begin{equation}
\hat{\lambda}\mathrm{(\Nh)} \propto \Nh^\mu,
\end{equation}
where $\hat{\lambda}\mathrm{(\Nh)}$ is the estimate of the surface density of Class~0/I YSOs at a column density \Nh, and $\mu$ is some power.
{\renewcommand{\labelenumi}{(\roman{enumi})}
\begin{enumerate}
\item It was found that four of the regions had Class~0/I populations inconsistent with $\mu = 0$ when combined with a threshold column density of $6\times10^{21} \Nh \mathrm{cm}^{-2}$ and zero probability elsewhere -- implying that star formation is not decoupled from column density (Section \ref{subsubsec:results-csr}).

\item The Class~0/I YSOs in all of the tested regions were also found to be inconsistent with $\mu=1$ -- the power law associated with the surface densities of prestellar cores (Section \ref{subsubsec:results-mu1}).

\item The power law index $\mu$ was measured for each region individually in Section \ref{subsec:mu}, the results of which are tabulated in Table \ref{table:mu}, and by combining the YSO surface density data from all regions a global $\mu$ value was measured to be $2.05 \pm 0.20$ where the reported uncertainty is the 95 per cent confidence interval.

\item The best value of $\mu$ tested was that of the global $\mu$ value 2.05 in Section \ref{subsubsec:results-mu2}, with only the YSOs in Serpens South and NGC1333 rejecting the model between 0.12~pc and 0.18~pc.
It was shown that all five regions were consistent with $\mu = 2.05$ when considering radial separations greater than 0.18~pc.

\item Serpens South and NGC1333 rejected the $\mu=2.05$ model at a radial separation of $\sim 0.15~\mathrm{pc}$. 
This could be due to physical effects such as a preferential scaling for filament collapse or small-scale interactions between YSO or data-related issues, such as resolution. 
However, because of the generally good fit to the model any modification should be limited to small spatial scale interactions.

\item Class~0/I YSOs were shown to have a different relationship to column density than Class~II YSOs (Section \ref{subsec:results-classII}) showing that this relationship is not consistent over time.

\item In Section \ref{subsec:disc-mu1}, using a toy evolution model it was determined that, 
if prestellar cores and protostars have different power-law relationships with column density, column density must play a role in their evolutionary timescales. Specifically, at least two of the prestellar core formation rate, prestellar core evolutionary time-scale and Class~0/I evolutionary time-scale, must be affected by the local column density environment. 

\end{enumerate}}

\begin{table}
\caption{Table of Symbols}
\label{table:symbols}
\begin{tabular}{ll}
\hline
Symbol & Description\\
\hline
$\Sigma_\mathrm{SFR}$ & star formation rate surface density \\
$\Sigma_\mathrm{GAS}$ & gas surface density \\
$\Cr$ & region-specific constant\\
$\mu$ & power-law index of SFR surface density relation \\
$\lambda$ & first-order intensity \\
$\Nh$ & column density $\mathrm{cm}^{-2}$\\
$N$ & number of YSOs \\
$u$,$v$ & cell indices \\
$\Delta \sigma$ & angular distance \\
$\alpha$ & right ascension (RA), significance level \\
$\delta$ & declination (Dec) \\
$r$ & radius, radial separation \\
$w$ & annulus width \\
$H_0$ & null hypothesis\\
$n$ & number of simulated patterns \\
$A$ & area \\
\hline
\end{tabular}
\end{table}

\section*{ACKNOWLEDGEMENTS}
Brendan Retter is funded by an STFC studentship. 
This research has made use of the {\sc starlink} software \citep{2014ASPC..485..391C} which is supported by the East Asian Observatory. 
The figures in this paper have been produced using {\sc matplotlib}: a 2D graphics package used in {\sc Python} for application development, interactive scripting, and publication-quality image generation across user interfaces and operating systems \citep{Hunter:2007}. 
This research made use of {\sc Astropy},\footnote{http://www.astropy.org} a community-developed core {\sc Python} package for Astronomy \citep{astropy:2013, astropy:2018}.
This research has made use of NASA’s Astrophysics Data System.
This work is based (in part) on observations made with the Spitzer Space Telescope, which is operated by the Jet Propulsion Laboratory, California Institute of Technology under a contract with NASA. 
This research has made use of data from the Herschel Gould Belt survey (HGBS) project (http://gouldbelt-herschel.cea.fr). The HGBS is a Herschel Key Programme jointly carried out by SPIRE Specialist Astronomy Group 3 (SAG 3), scientists of several institutes in the PACS Consortium (CEA Saclay, INAF-IFSI Rome and INAF-Arcetri, KU Leuven, MPIA Heidelberg), and scientists of the Herschel Science Center (HSC).

\section*{DATA AVAILABILITY}
The \textit{Herschel} Gould Belt survey (HGBS) data available in HGBS Archive, at \url{http://www.herschel.fr/cea/gouldbelt/en/} . 
The \citet{dunham2015} Young Stellar Object source data are available at \url{https://doi.org/10.1088/0067-0049/220/1/11} .
The Spitzer data underlying this article are available in NASA/IPAC Infrared Science Archive at \url{https://irsa.ipac.caltech.edu/data/SPITZER/C2D/images/}




\bibliographystyle{mnras}
\bibliography{YSOreading} 


\bsp	
\label{lastpage}
\end{document}